\documentclass[aps,prb,twocolumn,superscriptaddress,floatfix]{revtex4-1}
\usepackage{amsfonts,bm}
\usepackage{bbold}
\usepackage{amsmath}
\usepackage{amssymb}
\usepackage{amsthm}
\usepackage{graphicx}
\usepackage{graphics}
\usepackage{subfig}
\usepackage{color}
\usepackage{bm}
\usepackage{hyperref}
\usepackage{rotating}
\usepackage{comment}
\usepackage{caption}
\usepackage{indentfirst}
\usepackage[normalem]{ulem}
\hypersetup{
    colorlinks,
    citecolor=red,
    filecolor=black,
    linkcolor=black,
    urlcolor=black
}

\newcommand{\be}{\begin{equation} }
\newcommand{\ee}{\end{equation} }
\newcommand{\ba}{\begin{eqnarray} }
\newcommand{\ea}{\end{eqnarray} }

\newcommand{\bpm}{\begin{pmatrix}}
\newcommand{\epm}{\end{pmatrix}}
\newcommand{\bmm}{\begin{matrix}}
\newcommand{\emm}{\end{matrix}}

\newcommand{\bea}{\begin{eqnarray}}
\newcommand{\eea}{\end{eqnarray}}

\newcommand{\beq}{\begin{equation} }
\newcommand{\eeq}{\end{equation} }

\newcommand{\Z}{{\mathbb{Z}}}

\newcommand{\cO}{{\mathcal{O}}}

\newcommand{\cB}{{\mathcal{B}}}
\newcommand{\cA}{{\mathcal{A}}}
\newcommand{\cC}{{\mathcal{C}}}
\newcommand{\cR}{{\mathcal{R}}}

\newcommand{\indf}{{\rm{ind}_f}}
\newcommand{\tcR}{{\tilde{\cR}}}
\newcommand{\Cl}{{\text{C}\ell}}

\newcommand{\bit}{\begin{itemize}}
\newcommand{\eit}{\end{itemize}}

\begin{document}

\title{Interacting invariants for Floquet phases of fermions in two dimensions}

\author{Lukasz Fidkowski}
\affiliation{Department of Physics and Astronomy, Stony Brook University, Stony Brook, NY 11794, USA}
\affiliation{Kavli Institute for Theoretical Physics, University of California, Santa Barbara, CA 93106, USA}

\author{Hoi Chun Po}
\affiliation{Department of Physics, University of California, Berkeley, CA 94720, USA}
\affiliation{Department of Physics, Harvard University, Cambridge MA 02138, USA}

\author{Andrew C. Potter}
\affiliation{Department of Physics, University of Texas at Austin, Austin, TX 78712, USA}

\author{Ashvin Vishwanath}
\affiliation{Department of Physics, University of California, Berkeley, CA 94720, USA}
\affiliation{Department of Physics, Harvard University, Cambridge MA 02138, USA}

\begin{abstract}

We construct a many-body quantized invariant that sharply distinguishes among two dimensional non-equilibrium driven phases of interacting fermions.  This is an interacting generalization of a band-structure Floquet quasi-energy winding number\cite{rudner2013anomalous}, and describes chiral pumping of quantum information along the edge.  In particular, our invariant sharply distinguishes between a trivial and anomalous Floquet Anderson insulator in the interacting, many-body localized setting.  It also applies more generally to models where only fermion parity is conserved, where it differentiates between trivial models and ones that pump Kitaev Majorana chains to the boundary, such as ones recently introduced in the context of emergent fermions arising from eigenstate $\Z_2$ topological order\cite{FractionalCF}.  We evaluate our invariant for the edge of such a system with eigenstate $\Z_2$ topological order, and show that it is necessarily nonzero when the Floquet unitary exchanges electric and magnetic excitations, proving a connection between bulk anyonic symmetry and edge chirality conjectured in Ref. \onlinecite{FractionalCF}.

\end{abstract}

\maketitle

\section{Introduction}

Recently it has been shown that new band structures, having no equilibrium analogues, can arise in periodically driven free fermion systems \cite{rudner2013anomalous,roy2016periodic,Carpentier15, Nathan15, Nathan2016magnetization}.  They are characterized by new winding number topological invariants arising from the $2 \pi$ periodic nature of the quasi-energy spectrum, and with the addition of bulk disorder can give rise to a new type of single particle `Floquet' Anderson insulator \cite{kitagawa2010topological, titum2016anomalous}.  The stability of these band structures to interactions is not clear, however.  While a priori it may seem that all distinctions between interacting Floquet systems should be rendered meaningless because such systems are expected to absorb energy from the drive and heat up to infinite temperature, it has recently been shown that many-body localization (MBL) \cite{nandkishore2015many} can provide a robust way to avoid this heating problem \cite{lazarides2015fate, abanin2016theory, ponte2015many}.  Thus MBL provides a natural setting to study interacting Floquet phases \cite{jiang2011majorana,von2016phaseII, khemani2015phase,von2016absolute, else2016floquet,von2016phaseI,else2016floquet, roy2016abelian,MBLSPT,harper2016stability,FractionalCF} of fermions beyond the level of band structure analysis.  In this work we classify such interacting two dimensional Floquet phases of fermions, providing a many-body invariant that sharply distinguishes among them in the MBL setting.

The interacting 2d fermionic Floquet phases we focus on are dynamic counterparts of integer quantum Hall states, having no bulk topological order.  Despite this superficial similarity, they are inherently dynamical phases, exhibiting novel properties such as quantized chiral transport of quantum information \cite{po2016chiral, harper2016stability} that have no equilibrium analogue.  They were studied in Ref. \onlinecite{FractionalCF}, and several of their properties elucidated, including their emergent role in a bulk-boundary correspondence for a dynamic bosonic phase with $\Z_2$ topological order.  The main contributions of the present work, which is meant to complement Ref. \onlinecite{FractionalCF}, are (1) a rigorous classification of these fermionic phases, based on our construction of a many-body index sharply distinguishing among their interacting 1d edges, and (2) a proof, based on this classification, of the bulk-boundary correspondence proposed in Ref. \onlinecite{FractionalCF}.

As in the case of bosonic Floquet MBL phases \cite{po2016chiral}, our basic strategy is to use the full set of bulk local conserved quantities of the Floquet unitary operator to effectively decouple the stroboscopic edge dynamics from the bulk.  The non-trivial nature of the 2d bulk phase is then reflected in an anomalous property of these edge dynamics: namely, while the stroboscopic edge evolution preserves locality, in the sense of taking local operators to nearby local operators, it cannot be generated by any continuous evolution of a truly 1d local Floquet Hamiltonian.  In other words, it is not a finite depth quantum circuit of local unitaries.  A prototypical example of such an anomalous 1d edge is the chiral translation by one site: despite being locality-preserving, such a translation is not a finite depth quantum circuit.  Such bosonic 1d locality-preserving operators were fully classified, modulo finite depth quantum circuits, in Ref. \onlinecite{gross2012index}, and Ref. \onlinecite{po2016chiral} leveraged this classification to define a quantized many-body `chiral unitary' index that distinguishes among bosonic Floquet MBL phases.  However, due to its inherently bosonic nature, this classification cannot be directly applied to the fermionic problem.

The principal technical result that underlies the conclusions in the present work is a full classification of fermionic 1d locality preserving unitaries, modulo finite depth circuits.  This classification can be expressed as a quantized index $\nu_f = \zeta \log \sqrt{2} + \log \frac{p}{q}$, where $\zeta = 0,1$ is a $\Z_2$ index, and $p$ and $q$ are positive integers.  The $\log \frac{p}{q}$ portion of this index is the same as that obtained in the bosonic classification of Ref. \onlinecite{gross2012index}, and indeed we will show that 2d fermionic Floquet MBL systems with such indices are equivalent to their bosonic counterparts, if the latter are built out of `fundamental' fermionic degrees of freedom.  On the other hand, the $\log \sqrt{2}$ portion of the index is inherently fermionic.  An example of a fermionic locality preserving unitary with index $\log \sqrt{2}$ is a Majorana translation, defined by $\gamma_{i} \rightarrow \gamma_{i+1}$ in a Majorana mode representation of a fermionic 1d chain.  We construct a microscopic 2d fermionic Floquet MBL model exhibiting such a Majorana translation edge mode, which represents an inherently fermionic dynamical phase whose physical property is that a Kitaev chain is pumped onto the edge during every Floquet cycle.  We furthermore give a simple physical construction, in the general interacting setting, of a $\Z_2$ edge index that measures $\zeta$.

The dynamics in the Majorana 2d fermionic Floquet model does not conserve $U(1)$ particle number.  This means that any particle number conserving fermionic realization must include Goldstone modes of a $U(1)$ symmetry breaking order parameter, which would be problematic for many body localizability.  On the other hand, $\Z_2$ fermions can also appear as emergent excitations in $\Z_2$ topologically ordered bosonic systems without any symmetries, such as the Kitaev toric code \cite{ToricCode}.  In this setting, a `gauged' version of the Majorana 2d fermionic model was constructed in Ref. \onlinecite{FractionalCF}, with underlying toric code topological order.  The Floquet unitary in the model of Ref. \onlinecite{FractionalCF} has the property that it exchanges the $e$ and $m$ toric code quasi-particles, and thus gives an example of a Floquet enriched topological order (FET) \cite{potter2016dynamically}, in that the Floquet unitary acts as an anyonic symmetry.  Furthermore, Ref. \onlinecite{FractionalCF} shows that this model always exhibits a chiral edge mode, whose chiral unitary index is half that of a fundamental bosonic edge translation, and proposed that the bulk anyon-exchanging FET order is necessarily tied to such a fractional edge chirality.  On the other hand, it is certainly possible to have a global $\Z_2$ symmetry that exchanges $e$ and $m$ excitations in ordinary equilibrium toric code models, with no chiral edge modes.  Indeed, such models can be built out of commuting projectors, with the $\Z_2$ symmetry acting onsite \cite{Heinrich2016symmetry, Tarantino2016discrete}, which appears incompatible with the proposal of Ref. \onlinecite{FractionalCF}.  

We show that the resolution to this seeming paradox hinges on the difference between ground state topological order and eigenstate topological order.  To do this, we first give a precise strong definition of eigenstate topological order\cite{Chandran14,Bahri15,potter2016dynamically, Sid2016fractionalizing}, in terms of a set of localized $l$-bits that are equivalent to the vertex and plaquette terms of the standard square lattice toric code.  While demanding local unitary equivalence to the toric code is overly restrictive, since the model of interest might have a different geometry and different set of microscopic local degrees of freedom, we find that demanding stable local unitary equivalence, modulo trivial localized $l$-bit spins, gives a sufficiently robust definition of eigenstate topological order.  In particular, the honeycomb model of Ref. \onlinecite{FractionalCF} is stably equivalent to the square lattice toric code.

We will then focus on Floquet unitary evolutions that exchange the local conserved quantities corresponding to the charge ($e$) and flux ($m$).  As opposed to Floquet unitary evolutions that preserve all bulk local conserved quantities\cite{po2016chiral}, for which a bosonic edge state can be cleanly decoupled from the localized bulk, here we will see that no such decoupling is possible.  Nevertheless, we will see that it is still possible in this case to decouple an effective {\emph{fermionic}} edge.  We will then show, using our 1d fermionic classification, that this edge dynamics is stably equivalent to a Majorana translation -- and in particular is chiral -- precisely when the bulk Floquet unitary exchanges $e$ and $m$, proving the bulk FET - boundary chirality correspondence.  

\section{A $\Z_2$ invariant}

\subsection{Majorana SWAP model}

\begin{figure}[tb]
\begin{center}
\includegraphics[width=1\linewidth]{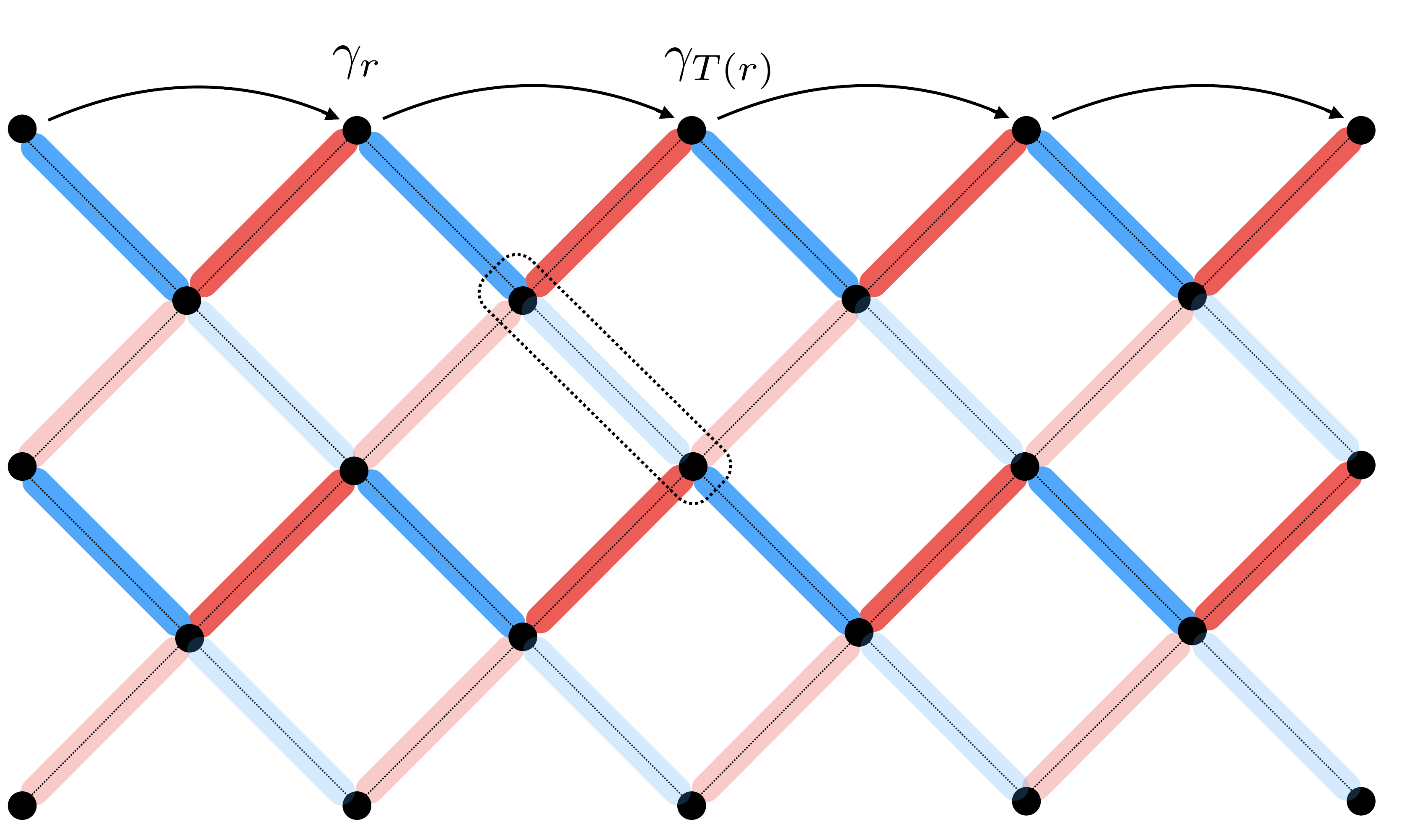}
\end{center}
\caption{Majorana SWAP model.}
\label{fig:fermion_fig1}
\end{figure}

\begin{figure}[tb]
\begin{center}
\includegraphics[width=1\linewidth]{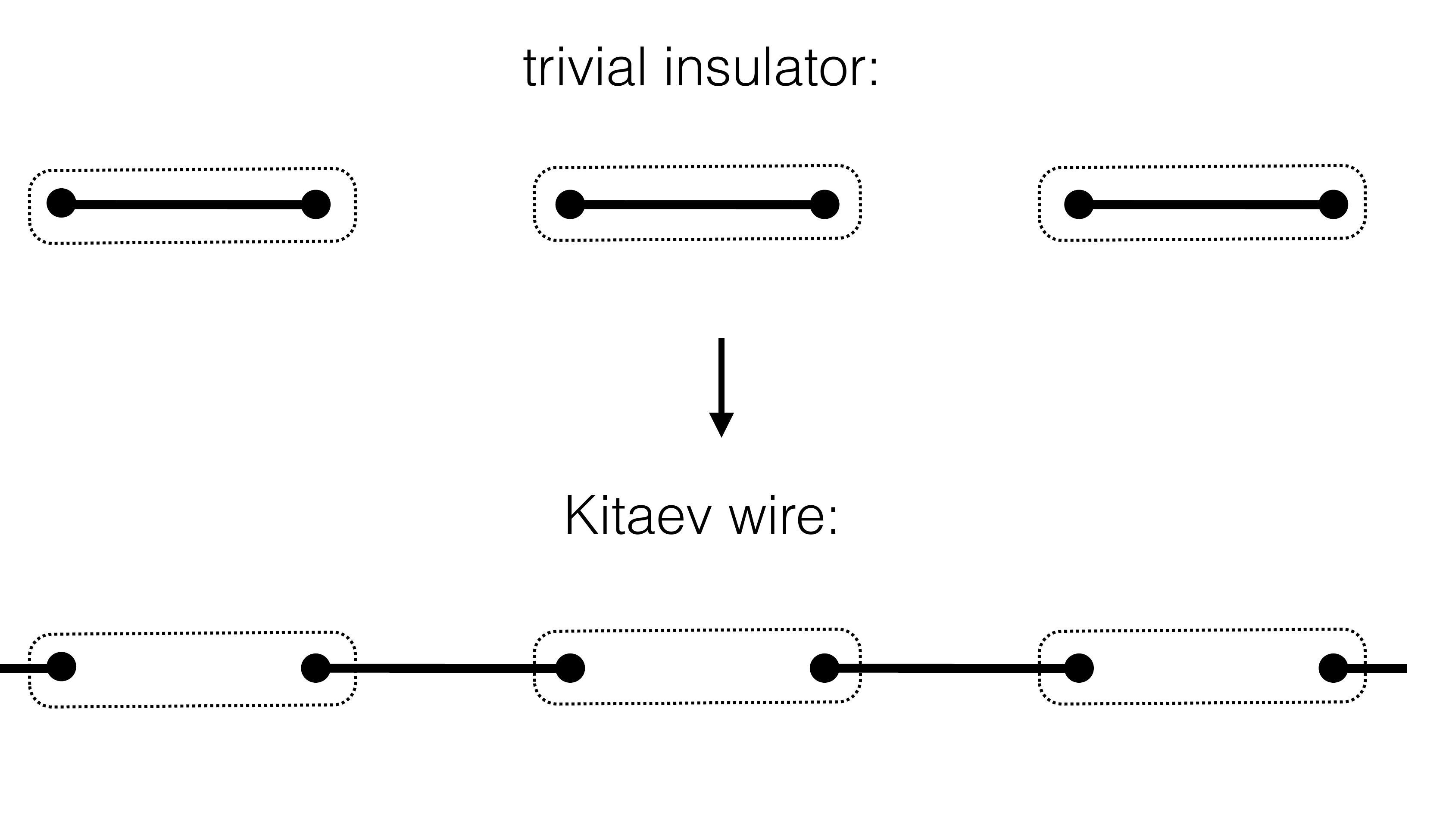}
\end{center}
\caption{The Floquet unitary acting on the boundary of the Majorana SWAP model pumps a Kitaev chain.}
\label{fig:fermion_fig2}
\end{figure}

The Majorana SWAP model is defined on a Hilbert space of Majorana zero modes $\gamma_r$ sitting on sites $r$ of a square lattice, as illustrated in Fig. \ref{fig:fermion_fig1}.  We will need an orientation on the links of this lattice.  We will pick it arbitrarily, and in our notation below always take a link $(r,r')$ to be oriented from $r$ to $r'$.  The Majorana modes $\gamma_r$ can be paired up into physical fermions by pairing sites, which we arbitrarily choose to be along the light blue links in Fig. \ref{fig:fermion_fig1}.  So for a light blue link $(r,r')$ we let
\begin{align}
a_{r,r'} &= \frac{1}{2} \left( \gamma_r + i \gamma_{r'} \right) \\
a_{r,r'}^\dag &= \frac{1}{2} \left( \gamma_r - i \gamma_{r'} \right)
\end{align}
so that the fermion parity of this physical fermion site is equal to
\begin{align}
{\cal P}_{r,r'} = 1-2 a_{r,r'}^\dag a_{r,r'} = i \gamma_r \gamma_{r'}
\end{align}

The Majorana SWAP Hamiltonian, periodic with period $T$, consists of 5 piecewise constant driving terms $H_j$, $j=1,\ldots, 5$, turned on for time $\frac{T}{5}$.  The first four 
\begin{align}
H_j = \sum_{(r,r') \in j} \frac{5\pi}{2T} ( i \gamma_r \gamma_{r'})
\end{align}
perform nearest-neighbor hops by turning on the solid blue, solid red, light blue, and light red links for $j=1,2,3,4$ respectively, as shown in Fig. \ref{fig:fermion_fig1}.  The fifth one
\begin{align}
H_5 = \sum_{(r,r')\in 3} W_{(r,r')} i \gamma_r \gamma_{r'}
\end{align}
is an onsite disorder term that is included for stability purposes.  Here the coupling constants $W_{(r,r')}$ are drawn from a uniform random distribution in $[-\frac{5 \pi}{T}, \frac{5 \pi}{T}]$.  We then see that the Floquet operator
\begin{align}
U(T)= \mathcal T \exp \left( -i \int_0^T dt \, H(t) \right)
\end{align}
is given by
\begin{align}
U(T)=U_5 U_4 U_3 U_2 U_1
\end{align}
where for $j=1,\ldots 4$,
\begin{align}
U_j= \prod_{(r,r') \in j} \exp \left( \frac{\pi}{2} \gamma_{r} \gamma_{r'} \right)
\end{align}
and
\begin{align}
U_5 = \prod_{(r,r') \in 3} \exp \left( \frac{T}{5} W_{(r,r')} \gamma_r \gamma_{r'} \right)
\end{align}

Under the $j$'th time step, the operators $\gamma_r, \gamma_{r'}$ in a link $(r,r')$ of color $j$ transform as
\begin{align}
\gamma_r &\rightarrow U_j \gamma_r U_j^\dag = - \gamma_{r'} \\
\gamma_{r'} &\rightarrow U_j \gamma_{r'} U_j^\dag = \gamma_r
\end{align}
so all of the $\gamma_r$ are invariant under the first 4 time steps:
\begin{align}
\gamma_r \rightarrow U \gamma_r U^\dag = \gamma_r.
\end{align}
and so in the bulk
\begin{align}
U(T) = U_5
\end{align}
Thus $\{ {\cal P}_{r,r'} \}$ for light blue links $(r,r')$ forms a full set of commuting local conserved quantities in the bulk of the system.

At the boundary of the system, the same analysis as in Refs. \onlinecite{rudner2013anomalous,po2016chiral}  shows that
\begin{align}
\gamma_r \rightarrow U(T) \gamma_r U(T)^\dag  = \gamma_{T(r)}
\end{align}
where $T(r)$ is a translation by one Majorana site, as indicated in Fig. \ref{fig:fermion_fig1}.  Acting on a trivial ground state of the effective 1d system, this Floquet unitary pumps a Majorana wire, as indicated in Fig. \ref{fig:fermion_fig2}.

\subsection{Decoupling edge and bulk in 2d fermionic system}

We will now show how to extract, for any 2d fermionic Floquet MBL system, a quasi 1d locality preserving fermionic unitary that describes the edge dynamics.  This discussion is similar to the one given in Ref. \onlinecite{po2016chiral} for bosonic systems.  We will then use the quasi 1d unitary to define a quantized many-body invariant that distinguishes the Majorana SWAP model constructed above from a trivial system.  Subsequently we will define a finer invariant which completely classifies all such quasi 1d unitaries, and thus gives a classification of fermionic 2d chiral Floquet MBL phases.

Take a lattice system with fundamental fermion degrees of freedom.  We will consider a general interacting local time-dependent Hamiltonian $H(t)$, periodic with period $T$, that conserves fermion parity.  Our MBL assumption then amounts to the existence of a full set of commuting local operators (FSCLO) that is conserved by the Floquet unitary $U(T)$.  This is just a set of commuting local operators with the property that specifying all of their eigenvalues fixes a state uniquely.  We will further assume that this FSCLO is adiabatically connected via a finite depth circuit of local unitaries $V$ to a set of trivial decoupled fermionic $l$-bits, i.e. a full set of conserved quantities on decoupled fermionic sites.

\begin{figure}[tb]
\begin{center}
\includegraphics[width=1\linewidth]{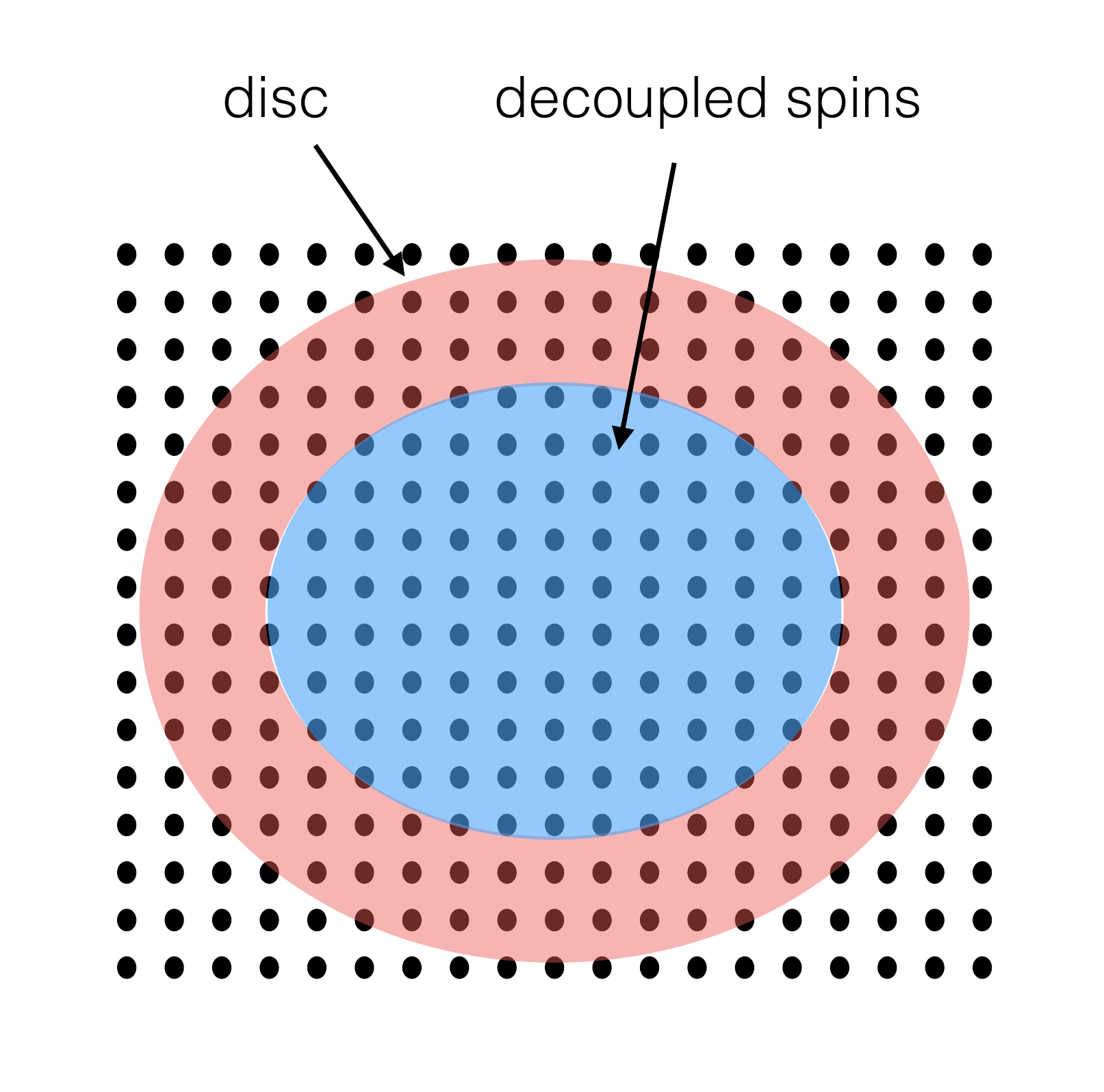}
\end{center}
\caption{After truncating to the disc and conjugating into the $l$-bit basis, the Floquet unitary preserves all of the spins in the bulk of the disc (blue).  Thus these spins can be set to arbitrary fixed values, resulting in a quasi 1d unitary operator $Y$ acting on the degrees of freedom near the edge (red).  The thickness of the edge must generally be taken to be greater than the Lieb-Robinson lengths $\xi$, $\xi'$ defined in the text.}
\label{fig:fermion_fig3}
\end{figure}

To extract the edge, it is first useful to discuss some of the length scales involved.  Besides the microscopic lattice spacing, there is a so-called `Lieb-Robinson' length $\xi$ associated to the Floquet unitary $U(T)$.  This is roughly equal to $T$ times the maximum of the Lieb-Robinson velocity of $H(t)$, and has the interpretation of a smearing length: if $X$ has support on some set of sites $S$, then $U(T)^\dag X U(T)$ will have most of its support on a `thickening' $S_{\xi}$ of $S$, consisting of all sites within distance $\xi$ of $S$.  In the following, we will assume that $U(T)$ has no exponential tails, i.e. $U(T)^\dag X U(T)$ is exactly supported on $S_{\xi}$.  This amounts to approximating $U(T)$ by a finite depth circuit; all of the arguments we give can be generalized from the finite depth circuit context to the general Floquet unitary context.  Similarly, there is a `Lieb-Robinson' length $\xi'$ associated to the finite depth circuit $V$. \footnote{The Lieb-Robinson length may be an overly cautious bound for MBL system, where the relevant length scale is the localization length which could in principle be much smaller.  Nevertheless, the Lieb-Robinson bounds suffice for the arguments made in the present paper.}

Now let
\begin{align}
U_{\text{disc}}(T) = \mathcal T \exp \left( -i \int_0^T dt \, H_{\text{disc}}(t) \right)
\end{align}
be the Floquet unitary for the Hamiltonian truncated to a large disc region (larger than $\xi$ or $\xi'$), denoted $H_{\text{disc}}(t)$.  Then deep in the bulk of the disc, at distances larger than $\xi$ from the edge, $U_{\text{disc}}(T)$ is the same as $U(T)$, and has a full set of bulk conserved quantities.

Now  let $V_{\text{disc}}$ denote an arbitrary truncation of $V$ to the disc, and define
\begin{align} \label{eq:trunc}
U'_{\text{disc}}(T) = V^\dag U_{\text{disc}}(T) V
\end{align}
Then deep in the bulk of the disc, at distances larger than $\xi$ or $\xi'$ from the edge, $U'_{\text{disc}}$ not only has a full set of conserved quantities, but these conserved quantities are simply decoupled fermionic sites.  We can thus restrict the evolution to a constrained Hilbert space where these bulk conserved quantities all have definite eigenvalues.  This then defines an effective fermionic locality preserving unitary $Y$ on the remaining degrees of freedom, which consist of sites near the edge.  This is illustrated in Fig. \ref{fig:fermion_fig3}. One can check that, up to deformation by finite depth circuits, $Y$ is independent of the choices made in this procedure.  In the next section we will define a $\Z_2$-valued many-body quantized index associated to $Y$ that distinguishes between the Majorana SWAP model and a trivial phase.

\subsection{A $\Z_2$-valued many body quantized invariant}

We now define a $\Z_2$-valued many-body invariant $\zeta \in \{0,1\}$ that separates 1d fermionic locality preserving unitaries -- and hence 2d fermionic chiral Floquet MBL phases -- into two distinct classes.  Both the trivial insulator and the anomalous Anderson Floquet insulator (AFAI)\cite{kitagawa2010topological, titum2016anomalous} are in the trivial class $\zeta=0$, whereas the Majorana SWAP model defined above has $\zeta=1$.  

The invariant $\zeta$ is defined as follows: given a locality preserving unitary $Y$, take a long interval $I$, and consider the evolved operator $Y^\dag P_I Y$.  Because $Y$ is locality preserving and fermion parity even, we expect that $Y^\dag P_I Y$ is equal to $P_I$ in the bulk of the interval $I$, i.e. the mismatch between $P_I$ and $Y^\dag P_I Y$ occurs only near the endpoints of $I$.  Formally, we expect that

\begin{align}\label{eq:Teq}
Y^\dag P_I Y  = P_I A_L A_R
\end{align}
where $A_L$ and $A_R$ are {\emph{local}} operators acting only on the sites near the left and right endpoints of $I$ respectively.  For a proof of Eq. \ref{eq:Teq}, see appendix \ref{app:ferm1}.

Now, from Eq. \ref{eq:Teq}, $A_L$ and $A_R$ must both have well defined fermionic parity, and their product must be fermion parity even.  Thus there are two possibilities: either $A_L$ and $A_R$ are either both fermion parity even, in which case we set $\zeta=0$, or they are both fermion parity odd, $\zeta=1$.

Clearly a trivial 2d insulator, for which $Y$ acts as the identity, has $\zeta=0$.  The AFAI also has $\zeta=0$.  Indeed, in this case the edge unitary performs a translation by a single fermionic site, so that the mismatch between $P_I$ and $Y^\dag P_I Y$ is given by operators at the left and right endpoints which measure the fermion parity of a single site; both are even operators.

For an example of a system with non-trivial $\zeta$, take the Majorana translation found at the edge of the Majorana SWAP model.  Formally, this edge can be described as follows.  Consider a periodic spinless fermion chain of length $N$, with creation and annihilation operators $a_n$ and $a^\dag_n$ at site $n$, and re-write these in terms of $2N$ Majorana modes:

\begin{align}
a_n &= \frac{1}{2} (\gamma_{2n-1} + i \gamma_{2n})\\
a^\dag_n &= \frac{1}{2} (\gamma_{2n-1} - i \gamma_{2n})
\end{align}
Then $Y$ acts by
\begin{align}
Y^\dag \gamma_i Y = \gamma_{i+1}
\end{align}
for $i=1,\ldots, 2N-1$, and $U^\dag \gamma_{2N} U = - \gamma_1$.  Explicitly, $Y$ can be constructed as a unitary operator as follows.  Let $M$ be the $2N$ by $2N$ matrix defined by $M_{i,i+1}=1$ for $i=1,\ldots, 2N-1$, $M_{2N,1}=-1$, and all other $M_{i,j}=0$.  Since $M$ is in $SO(2N)$, it can be written as $M=\exp(A)$, with $A$ real and anti-symmetric.  Then letting
\begin{align} \label{Maj}
Y=\exp \left( \frac{1}{4} \sum_{i,j} A_{i,j} \gamma_i \gamma_j \right)
\end{align}
we see that $Y$ acts on the $\gamma_i$ as desired.  Note that $A$ is not local, in the sense that it has non-zero matrix elements $A_{i,j}$ for large $|i-j|$, so that $Y$ is not a finite depth quantum circuit; nevertheless, it is locality-preserving.

To see that the Majorana translation has $\zeta=1$, write the fermion parity $P_I$ of an interval $I=[a,b]$ as
\begin{align}
P_I = (i \gamma_{2a-1} \gamma_{2a}) (i \gamma_{2a+1} \gamma_{2a+2}) \ldots (i \gamma_{2b-1} \gamma_{2b}).
\end{align}
Then
\begin{align}
Y^\dag P_I Y = i^{b-a+1} \gamma_{2a} \gamma_{2a+1} \ldots \gamma_{2b+1}
\end{align}
so that the mismatch defined in Eq. \ref{eq:Teq} is, up to sign, $A_L=\gamma_{2a-1}$ and $A_R = \gamma_{2b+1}$.  These are both fermion parity odd operators, and hence $\zeta=1$.

\section{Classification of 1d fermionic locality preserving unitaries and 2d fermionic Floquet MBL phases}

In the previous section we defined a $\Z_2$-valued many-body invariant $\zeta$ that gives a coarse classification of 2d fermionic Floquet MBL systems into those that pump a Kitaev chain to the boundary and those that do not.  However, this invariant does not distinguish between a trivial phase and the anomalous Floquet Anderson insulator (AFAI), which performs a chiral translation by a physical fermionic site at the edge.  Although a many-body rational-valued index that distinguishes among bosonic analogues of the AFAI has been defined\cite{po2016chiral, gross2012index}, it is not a priori clear whether it remains stable or becomes modified in the presence of fermionic degrees of freedom.

In this section, we define such an index $\nu_f(Y)$ in the fermionic setting.  Let us first say precisely what we mean by `fermionic setting'.  On the one hand, we could take this to mean systems whose site Hilbert spaces are generated by some number $2n$ of Majorana zero modes -- i.e. they are $2^n$ dimensional Fermionic Fock spaces based on $n$ physical fermionic modes.  In this case we will show in subsection \ref{subsec:restricted} below that $\nu_f(Y)$ takes the form
\begin{align}
\nu_f(Y) = \frac{k}{2} \log 2,
\end{align}
where $k$ is an integer that describes the chiral nature of $Y$.  Specifically, non-zero even $k$ corresponds to a translation by some number of physical fermion sites, as in the AFAI edge, whereas odd $k$ corresponds to a net Majorana translation; in particular $\zeta=k\,\text{mod}\,2$.  More generally, however, we want to consider a class of systems that allows for both fermions and general bosonic systems.  This is the setting of general $\Z_2$-graded Hilbert spaces, which we treat in subsection \ref{subsec:general}.  In this case we show that $\nu_f(Y)$ takes the form
\begin{align}
\nu_f(Y) = \frac{\zeta}{2} \log 2 + \log \left(\frac{p}{q}\right)
\end{align}
where $p$ and $q$ are relatively prime positive integers whose prime factors are all divisors of the site Hilbert space dimensions.

The physical interpretation of $\nu_f(Y)$ is that it characterizes the extent to which $Y$, despite being locality preserving, cannot be generated as the Floquet evolution of any local 1d fermionic Floquet Hamiltonian.  The most important property of $\nu_f(Y)$ is that, as we discuss in subsection \ref{subsec:classification}, it is the only obstruction to the existence of such a 1d generating Hamiltonian.  Thus $\nu_f(Y)=1$ implies that $Y$ is a finite depth quantum circuit, and $\nu_f(Y)=\nu_f(Y')$ implies that $Y'$ differs from $Y$ by finite depth circuits $U$ and $W$: $Y'=UYW$.  There is one caveat, as we discuss in subsection \ref{subsec:classification}: in order for these equations to be true, we might have to allow for additional ancilla fermionic degrees of freedom, leading to a notion of stable equivalence.  The final result then is that fermionic systems with $\zeta=0$ are stably equivalent to bosonic systems (in particular, the AFAI is stably equivalent to the bosonic SWAP model of Ref. \onlinecite{po2016chiral}), all bosonic systems with non-trivial chiral unitary index remain non-trivial in the presence of fermions, and modulo these bosonic systems there is only one non-trivial fermionic equivalence class, namely that of the Majorana SWAP model introduced above.

\subsection{Definition of quantized many body index: Fermionic Fock space} \label{subsec:restricted}

Let us give a precise definition of the fermionic chiral unitary index $\nu_f(Y)$ for a fermionic locality-preserving unitary $Y$.  In this subsection we will restrict for simplicity to fermionic systems whose site Hilbert spaces are $2^n$ dimensional and can be thought of as a Fock space of $n$ independent fermionic modes.  In this setting we will show that $\nu_f=\frac{1}{2} \log 2$ in the case of a Majorana translation and that $\nu_f=\log 2$ in the case of the AFAI.


To start, take a large but finite system $\Lambda$ with periodic boundary conditions, with lattice sites labeled by $x$ at which sit fermionic Fock spaces ${\cal H}_x$ of dimension $2^{n_x}$.  Then the algebra ${\cal A}_x$ of local operators at site $x$ is $2^{2n_x}$ dimensional, and is generated by $2n_x$ Majorana modes $\gamma_{x,1},\ldots \gamma_{x,2n_x}$.  For a collection of sites $S$ we let ${\cal A}_S = \otimes_{x\in S} {\cal A}_x$ be the algebra of operators supported on $S$.

Now take a spatial cut and consider two contiguous intervals of sites $L$ and $R$, residing immediately to the left and right of the cut respectively.  We require that $L$ and $R$ are longer than the Lieb-Robinson length of $Y$.  The algebras ${\cal A}_L$ and ${\cal A}_R$ then commute in the $\Z_2$-graded sense.  This just means that fermion parity even operators of each algebra commute with everything in the other algebra, and fermion parity odd operators anti-commute.  On the other hand, $Y^\dag {\cal A}_L Y$ and ${\cal A}_R$ might fail to commute in this $\Z_2$-graded sense, which is an indication of a flow of quantum information from the left to the right.  To quantify this, we define a general measure $\eta({\cal A},{\cal B})$ that describes the extent to which algebras ${\cal A}$ and ${\cal B}$ fail to graded-commute.  Letting ${\cal A}$ and ${\cal B}$ be generated by $2 n_A$ and $2 n_B$ Majorana modes respectively, we can form monomials of these to generate sets of $p_A = 2^{2n_A}$ and $p_B=2^{2n_B}$ orthonormal operators $e^a_i, i=1,\ldots, p_A$ and $e^b_j, j=1,\ldots,p_B$ spanning ${\cal A}$ and ${\cal B}$ respectively.  We then use these to define
\begin{align}\label{eq:defeta}
\eta({\cal A},{\cal B}) = 2^{ - n_{\Lambda}} \sqrt{\sum_{i=1}^{p_A} \sum_{j=1}^{p_B}
|\text{Tr}_{\Lambda} \left({e^a_i}^\dag e^b_j \right)|^2}
\end{align}
where
\begin{align}
n_{\Lambda}=\sum_{x\in \Lambda} n_x
\end{align}
is the total number of fermionic modes in the whole system, and the trace in Eq. \ref{eq:defeta} is taken in the Hilbert space of the whole system.  Using this, we then define:
\begin{align} \label{eq:defnu}
\nu_f(Y) = \text{log} \left(\frac{\eta(Y^\dag {\cal A}_L Y, {\cal A}_R)}{\eta({\cal A}_L, Y^\dag {\cal A_R} Y)}\right)
\end{align}
Although it is not obvious from its definition in Eq. \ref{eq:defnu}, we demonstrate in the next subsection (see also appendix \ref{app:ferm1}) that $\nu_f(Y)$ is independent of the choices made in the definition, and takes the quantized form $\nu_f(Y) = \frac{n}{2} \log 2$.

Let us evaluate $\nu_f(Y)$ for the Majorana translation $\gamma_i \rightarrow \gamma_{i+1}$ in a chain of spinless fermions.  It suffices to take $L$ and $R$ to consist of a single fermionic mode each, described by $\gamma_1,\gamma_2$ and $\gamma_3,\gamma_4$ respectively.  Then it is clear that the denominator in the logarithm in Eq. \ref{eq:defnu} is equal to $1$, since the two algebras ${\cal A}_L$ and $Y^\dag {\cal A_R} Y$ graded-commute.  As for the numerator, there are exactly two non-zero contributions to the sum in Eq. \ref{eq:defeta}, coming from $e_i^a=e_j^b=1$ and $e_i^a=e_j^b=\gamma_3$.  Both contribute $2^{n_{\Lambda}}$ (the dimension of the total Hilbert space) to the trace, so that $\nu_f(Y)=\log \sqrt{2}$.  A similar argument shows that a translation by one physical site, such as that occuring at the edge of the AFAI, has a fermionic index equal to $\nu_f=\log 2$, and is hence non-trivial.

\subsection{Definition of quantized many body index: general $\Z_2$-graded Hilbert space}\label{subsec:general}

In order to treat fermionic and bosonic systems on the same footing, we now define the index in the more general setting of so-called $\Z_2$-graded site Hilbert spaces.  A $\Z_2$-graded Hilbert space $H_i$ is a Hilbert space that can be decomposed into fermion parity even and odd pieces
\begin{align}
H_i = H_i^0 \oplus H_i^1
\end{align}
Letting $p_i= |H_i^0|$ and $q_i=|H_i^1|$ be the dimensions of the odd and even sectors of $H_i$, we will also use the notation $H_i={\mathbb C}^{p|q}$.  The total Hilbert space $H$ is now the tensor product of the site Hilbert spaces $H_i$:
\begin{align} \label{defH}
H=\otimes_i H_i
\end{align}
$H$ also has a natural $\Z_2$-grading: the total fermion parity in $H$ is the product of the individual fermion parities in the $H_i$.

For example, a spinless fermion is described by a two dimensional $\Z_2$-graded site Hilbert space $H_i = {\mathbb C}^{1|1}$.  The total Hilbert space $H$ is then just the usual fermionic Fock space.  A spinful fermion can similarly be described by $H_i = {\mathbb C}^{2|2}$.  On the other hand, if we take $H_i$ to be purely even -- i.e. $H_i={\mathbb C}^{p|0}$ -- then we recover a purely bosonic system.  Our framework encompasses all of these cases.

Let us denote the algebra of all operators on $H_i$ by $\cO_i$: this is simply the algebra of $|p_i+q_i|$ by $|p_i+q_i|$ complex matrices.  We will also use the notation $\cO_i={\mathbb C} (p_i | q_i)$.  Again, $\cO_i$ splits into even and odd components:

\begin{align}
\cO_i = \cO_i^0 \oplus \cO_i^1
\end{align}
$\cO_i^0$ is the sub-algebra of all operators that conserve fermion parity, and is just ${\mathbb C}(p_i) \oplus {\mathbb C}(q_i)$.  Its dimension as a complex vector space is thus $p_i^2 + q_i^2$.  $\cO_i^1$ is the space operators that mix the two fermion parity sectors, and has dimension dimension $2 p_i q_i$.  We will say that an operator $X \in \cO_i^j$, $j=0,1$, has well defined fermion parity $j$, and set $|X|=j$.

More generally, for a set of sites $S$, we define the algebra of operators supported on $S$ as

\begin{align}
\cO_S = \otimes_{i \in S} \, {\cal O}_i
\end{align}
In the notation above $\otimes$ represents the $\Z_2$-graded tensor product, which just means that odd operators on distinct sites anti-commute.  By tensoring with the identity on all sites not in $S$, we can view each such subalgebra ${\cal O}_S$ as sitting inside the algebra of operators on all of $H$, which we simply denote ${\cal O}$.  We will then say that operators in ${\cal O}_S \subset {\cal O}$ are `supported' on $S$.

The anti-commuting nature of fermionic operators on distinct sites motivates the following definition of the graded commutator $[X,Y]_g$ of two operators $X,Y$ of well defined fermion parity:
\begin{align}
[X,Y]_g \equiv X Y - (-1)^{|X| |Y|} Y X,
\end{align}
By linearity the definition of graded commutator extends to all operators, not just those of well defined fermion parity.  Then $[\cO_i, \cO_j]_g=0$ on distinct sites $i,j$.

Given a locality-preserving operator $Y$, we now sketch an algebraic definition of the index $\nu_f(Y)$ in the general $\Z_2$-graded case, which makes it manifest that the index is quantized in the claimed form -- for more details, including why this definition is equivalent to the one in Eq. \ref{eq:defnu} for Fermionic Fock spaces, see appendix \ref{app:ferm1}.  First, coarse grain the Hilbert space by grouping sites in such a way that $Y$ is locality-preserving with range $1$.   Now take the operator algebra on two neighboring sites, $\cO_{2x} \otimes \cO_{2x+1}$.  It follows that

\begin{align}
Y^\dag & \left(\cO_{2x} \otimes \cO_{2x+1} \right) Y \\
&\subset  \left(\cO_{2x-1} \otimes \cO_{2x}\right) \otimes \left(\cO_{2x+1} \otimes \cO_{2x+2} \right)
\end{align}
We now want to quantify the extent to which $Y^\dag \left(\cO_{2x} \otimes \cO_{2x+1} \right) Y$ is supported on either of the two tensor factors in brackets on the right hand side of the above equation, which will reflect the chiral nature of $Y$.  In appendix \ref{app:ferm1} we show, using the fact that conjugation by $Y$ preserves the $\Z_2$-graded algebra structure, that there exist mutually graded-commuting `support' algebras
\begin{align}
{\cal R}_{2x} &\subset {\cal O}_{2x-1} \otimes {\cal O}_{2x} \\
{\cal R}_{2x+1} &\subset {\cal O}_{2x+1} \otimes {\cal O}_{2x+2}
\end{align}
such that
\begin{align}\label{eq:a}
Y^\dag \left(\cO_{2x} \otimes \cO_{2x+1} \right) Y &= {\cal R}_{2x} \otimes {\cal R}_{2x+1} \\
{\cal O}_{2x+1} \otimes {\cal O}_{2x+2} &= {\cal R}_{2x+1} \otimes {\cal R}_{2x+2}
\end{align}
The proof of Eq. \ref{eq:a}, given in appendix \ref{app:ferm1}, is non-trivial, and requires generalizing the algebraic constructions of Ref. \onlinecite{gross2012index} to the $\Z_2$-graded algebra setting.

Taking dimensions of both sides of Eq. \ref{eq:a} shows that
\begin{align}
&\frac{\sqrt{|\cR_{2x}|}}{|{\cal H}_{2x}|} = \frac{|{\cal H}_{2x+1}|}{\sqrt{|\cR_{2x+1}|}} \equiv \indf (Y)
\end{align}
is independent of $x$.  Furthermore, this equation shows that $(\indf(Y))^2$ is a rational number $p/q$ with all of the prime factors of $p$ and $q$ being divisors of the site Hilbert space dimensions.  However, one can say more.  As we show in appendix \ref{app:ferm1}, the algebras ${\cal R}_y$ are simple in the $\Z_2$ graded sense \cite{Fidkowski11,Bultinck2016fermionic}.  Such simple $\Z_2$ graded algebras come in precisely two forms: (1) even algebras, which are matrix algebras over a $\Z_2$-graded vector space and have dimension $d^2$, or (2) odd algebras, which are matrix algebras over an odd Clifford algebra and have dimension $2d^2$.  This shows that $\indf(Y)$ is either a rational number or a rational number times the square root of two.  $\nu_f(Y)$ is defined by taking its logarithm:

\begin{align}
\nu_f(Y)=\log \indf(Y)
\end{align}
Thus $\nu_f(Y)$ must be of the form
\begin{align}
\nu_f(Y) = \frac{\zeta}{2} \log 2 + \log \frac{p}{q}
\end{align}

\subsection{Properties of $\nu_f$} \label{subsec:classification}

{\bf Rational versus radical: } The fact that $\indf (U)$, defined in Eq. \ref{defind}, does not depend on $x$ implies that the $\cR_y$ are either all even simple $\Z_2$-graded algebras, or they are all odd.  In the former case $\indf (U)$ is a rational number -- we will refer to this case as `rational' ($\zeta=0$) -- and in the latter case it is a rational number times the square root of $2$ -- we will refer to this case as `radical' ($\zeta=1$).

\vspace{2mm}

{\bf Invariance under deformation by finite depth circuits: } We claim that $\indf(U)=\indf(VUV')$ for any finite depth circuits $V$ and $V'$.  Indeed, $VUV'$ can be continuously connected to $U$ in the space of locality-preserving unitaries, simply by continuously deforming $V$ and $V'$ to the identity.  Since $\indf$ is a continuous discrete valued function on locality-preserving unitaries, it must be constant on connected components, and hence $\indf(U)=\indf(VUV')$.

\vspace{2mm}

{\bf Multiplicativity: } The fermionic index also satisfies the property that, for two different locality-preserving unitaries $U,U'$,

\begin{align} \label{mult}
\indf (U U')=\indf(U) \indf(U')
\end{align}
Furthermore, stacking two disjoint systems with locality-preserving unitaries $U,U'$, we obtain
\begin{align} \label{stacking}
\indf(U \otimes U')=\indf(U) \indf(U')
\end{align}
The proof of Eq. \ref{stacking} follows directly from the formula \ref{defind}.  Since $U \otimes U'$ can be smoothly connected to $UU' \otimes 1$, Eq. \ref{mult} follows from the fact that $\indf$ is locally constant.

Taking logarithms, the fermionic chiral unitary index satisfies corresponding additivity properties:
\begin{align}
\nu_f (U U') &= \nu_f (U) + \nu_f (U') \\
\nu_f (U \otimes U') &= \nu_f (U) + \nu_f (U')
\end{align}

{\bf Completeness of classification: } The most non-trivial property of $\indf$, and hence $\nu_f$, is that it completely classifies 1d fermionic locality-preserving unitaries.  Let us explain carefully what we mean by this, because the fermionic situation is somewhat more subtle than the bosonic one studied in Refs. \onlinecite{po2016chiral,gross2012index}.  In the bosonic situation, if two locality-preserving unitaries $Y$ and $Y'$ had the same chiral unitary index, then they were necessarily related by finite depth circuits $U,V$: $Y'=UYV$.  However, in the case of fermions this is not true.  For example, consider stacking a bosonic spin-1/2 system, with site Hilbert spaces ${\mathbb C^{2|0}}$, on top of a spinless fermion system, with site Hilbert spaces ${\mathbb C^{1|1}}$, so that the total system has site Hilbert spaces ${\mathbb C^{2|0}} \otimes {\mathbb C^{1|1}} = {\mathbb C^{2|2}}$.  Now, a translation by one site in the bosonic subsystem turns out to be not deformable to a translation by one site in the fermionic subsystem, even though the two have the same $\nu_f=2$.

To claim that $\nu_f$ gives a complete classification in the fermionic case will thus require a more general notion of equivalence.  To this end, two fermionic locality preserving unitaries $Y$ and $Y'$ on the same Hilbert space are said to be {\emph{stably equivalent}} if upon appropriately enlarging the Hilbert space by appending inert ancilla fermionic degrees of freedom, one can find finite depth circuits $U,V$ in this larger Hilbert space such that:
\begin{align}
Y' \otimes {\mathbb 1}_f = U \left( Y \otimes {\mathbb 1}_f \right) V
\end{align}
Then we prove in appendix \ref{app:ferm1} that:
\vspace{2mm}

{\bf Claim: } If $\nu_f (Y)=\nu_f(Y')$, then $Y$ and $Y'$ are stably equivalent.

\vspace{2mm}
The physical implications of this claim are as follows.  First, for any rational fermionic $Y$, which has $\nu_f(Y)= \log \frac{p}{q}$ for some integers $p,q$, we can find a bosonic system $Y'$ with the same value of the chiral unitary index.  If we consider this bosonic system in the setting of $\Z_2$ graded Hilbert spaces, then the result above implies that $Y$ and $Y'$ are stably equivalent.  Thus, all rational fermionic locality-preserving unitaries are stably equivalent to bosonic ones.  Furthermore, modulo such bosonic locality-preserving unitaries, there is only one non-trivial fermionic locality-preserving unitary, namely the Majorana translation, with $\zeta=1$.

\section{Bulk-boundary correspondence for anyon permuting symmetry in the toric code}

One problematic feature of the Majorana SWAP model is that its driving Hamiltonian violates particle number conservation, conserving only the fermionic parity.  Thus a physical realization of this model in a particle number conserving setting will require a spontaneous breaking $U(1)$ symmetry breaking, leading to Goldstone modes, which are problematic for MBL.  On the other hand, fermions can also arise as emergent excitations in a bosonic model with topological order, and arbitrary fermion parity conserving interactions can be engineered in this setting.  Ref. \onlinecite{FractionalCF} uses this strategy in the context of the spin-$1/2$ Honeycomb model \cite{kitaev2006anyons} to design a driving Hamiltonian that is effectively a fermion-parity gauged version of the Majorana SWAP model.

The model of Ref. \onlinecite{FractionalCF} has two interesting properties.  First, the Floquet unitary exchanges $e$ and $m$ excitations for all eigenstates.  Second, it has a chiral edge.  Because the $e$ and $m$ excitations are exchanged rather than truly conserved in the bulk, one cannot decouple an edge directly.  Rather, Ref. \onlinecite{FractionalCF} shows that $U(2T)$, which does have a full set of conserved bulk quantities, performs a chiral translation at the edge, and shows that its chiral unitary index is $\log 2$.  This means that, insofar as an edge for $U(T)$ could be decoupled, it would have a fractional index $\frac{1}{2} \log 2$.  More generally, Ref. \onlinecite{FractionalCF} proposes that such a fractional chiral edge occurs for any Floquet evolution that exchanges $e$ and $m$ excitations in a system with eigenstate topological order.

In the rest of this section, we use the fermionic machinery developed above to prove this correspondence.  We first give a precise `strong' definition of eigenstate topological order, in terms of stable equivalence to toric code projectors.  This definition is strong in the sense that any system that satisfies it will also have eigenstate topological order\cite{Chandran14,Bahri15,potter2016dynamically} according to any other definition, such as one that uses the existence of string operators that commute with the Hamiltonian.  However, this strong definition is also sufficiently robust to capture the honeycomb Majorana SWAP model, as we show in appendix \ref{app:stable}.  We conjecture that it is actually equivalent to other definitions of eigenstate topological order, but leave this for future work.  We then show that although there is no way to decouple a bosonic edge for a Floquet unitary $U(T)$ that exchanges $e$ and $m$ excitations in a system with eigenstate topological order, one can decouple a well defined fermionic edge.  Using the fermionic machinery developed above we then show that this fermionic edge has $\zeta=1$, i.e. is in the radical class, which proves the bulk-boundary correspondence.

It should be noted that it is possible to exchange $e$ and $m$ excitations in a model with ground state topological order using a purely onsite unitary operator\cite{Heinrich2016symmetry, Tarantino2016discrete, meng2016exactly}.  However, our proof makes explicit use of topological order in all eigenstates, not just the ground state, in the form of the `strong' eigenstate topological order mentioned above.  This is the essential difference between our setting and the ground state situation: although the models of Refs. \onlinecite{Heinrich2016symmetry, Tarantino2016discrete, meng2016exactly} are built out of commuting projectors, these commuting projectors do not lead to eigenstate topological order in the `strong' sense.  Indeed, these models are built in such a way that the only non-trivial dynamics occurs close to the ground state, leading to large degeneracies in excited states, in opposition to the case of eigenstate topological order.

\subsection{Eigenstate topological order and anyon-permuting Floquet evolutions} \label{subsec:eig}

\begin{figure}[tb]
\begin{center}
\includegraphics[width=1\linewidth]{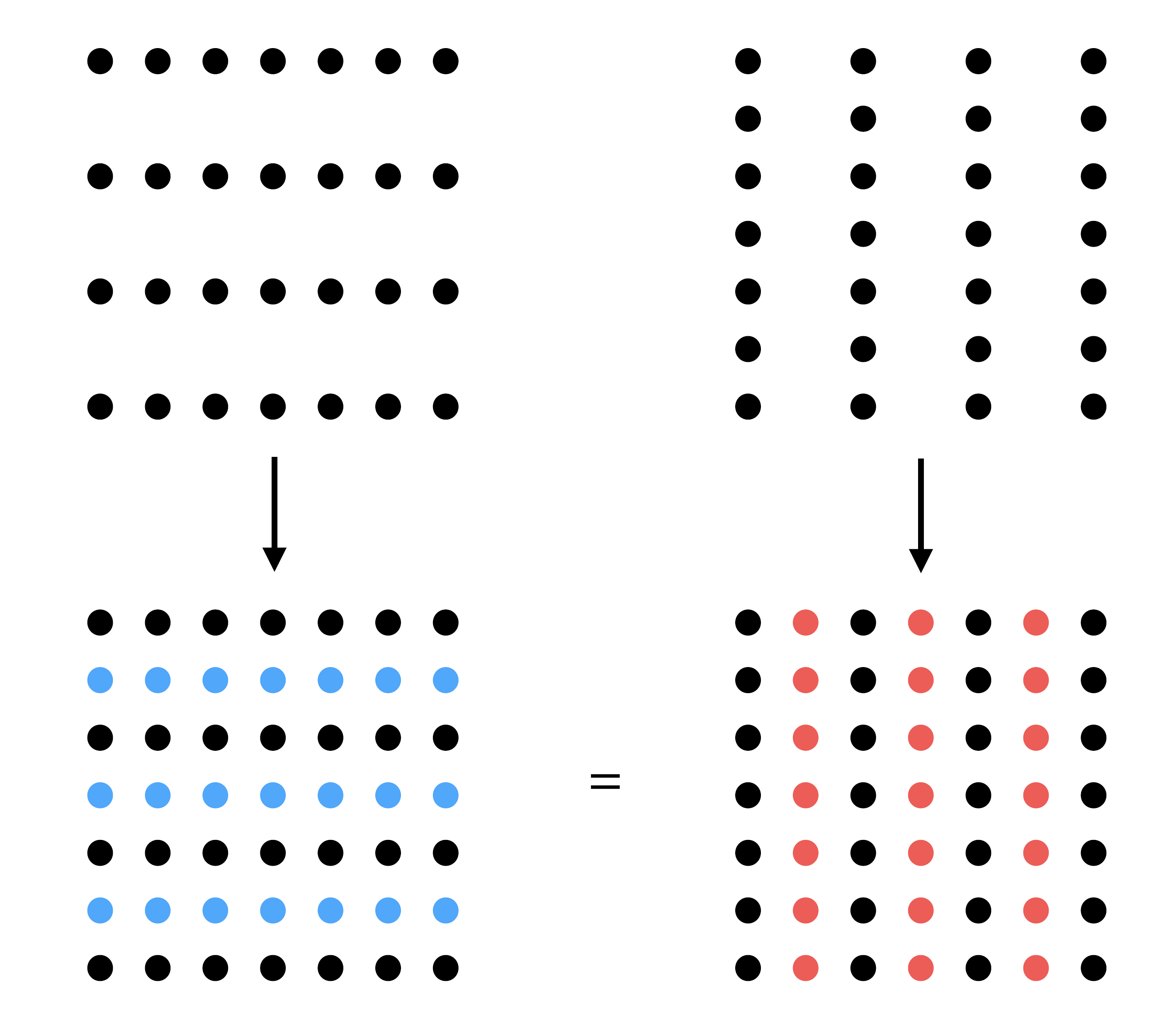}
\end{center}
\caption{The spin-$1/2$ systems in the first row are built on different microscopic Hilbert spaces.  However, after the addition of appropriate ancilla spins in each case (blue and red respectively), the microscopic Hilbert spaces become equivalent.}
\label{fig:stable_eq_fig7}
\end{figure}

Consider two many-body Hilbert spaces 1 and 2, each with a full set of commuting local operators (FSCLO) $\{  {\cal O}_j^{(1)} \}$ and $\{  {\cal O}_j^{(2)} \}$ respectively.  By full set we mean simply that specifying all their eigenvalues determines a state uniquely \footnote{we allow topological degeneracies on surfaces of non-trivial topology}.  We do not necessarily demand that these many-body Hilbert spaces are built on identical microscopic degrees of freedom.  However, suppose that it is possible to add ancilla local degrees of freedom -- e.g. spin $1/2$'s -- to both many-body Hilbert spaces, such that the resulting enlarged Hilbert spaces do have identical microscopic degrees of freedom (see Fig \ref{fig:stable_eq_fig7}).  Enlarging the FSCLO's by appending $\sigma^z$ operators for all of the ancilla spins results in FSCLOs $\{   {{\cal O}'}_j^{(1)} \}$ and $\{  {{\cal O}'}_j^{(2)} \}$ for the enlarged Hilbert space.  We then say that the original FSCLOs for system 1 and system 2 are stably equivalent if there exists a finite depth circuit of local unitaries $V$ in the enlarged Hilbert space that maps the commuting algebra generated by $\{   {{\cal O}'}_j^{(1)} \}$ to the commuting algebra generated by $\{  {{\cal O}'}_j^{(2)} \}$.  This definition can readily be generalized to the case of ancilla degrees of freedom with more general site Hilbert space dimension $p>2$.

In appendix \ref{app:stable} we construct a honeycomb model which is a slight variant of that constructed in Ref. \onlinecite{FractionalCF}, and show that its FSCLO is stably equivalent to the usual square lattice toric code projectors.  These are the standard vertex and plaquette terms ${\cal A}_V$ and ${\cal B}_F$, associated to vertices $V$ and faces $F$ of the toric code square lattice:

\begin{align}
{\cal A}_V &= \prod_{l \sim V} S_{l}^x \\
{\cal B}_F &= \prod_{l \in \partial F} S_{l}^z,
\end{align}
where $l$ labels the links of the square toric code lattice.  We will use stable equivalence to the toric code as our `strong' definition of $\Z_2$ eigenstate topological order.  In particular, as shown in appendix \ref{app:stable}, the honeycomb model conserved quantities used in Ref. \onlinecite{FractionalCF} are stably equivalent to the square lattice toric code, and hence exhibit `strong' eigenstate order in our sense.

Assuming now that we have a system with a FSCLO $\{ {\cal O}_j \}$ that is stably equivalent -- via a finite depth circuit $U$ -- to the standard toric code, let us try to understand Floquet dynamics that is compatible with this FSCLO.  Demanding that all of the ${\cal O}_j$ are conserved under Floquet evolution -- i.e. $U_F^\dag {\cal O}_j U_F = {\cal O}_j$ -- is overly restrictive, because it rules out Floquet operators of the sort we want to study, namely ones that exchange $e$ and $m$.  Instead, we will demand the following weaker:

\vspace{5pt}

{\bf Compatibility condition:}  The local operators $\{U_F^\dag {\cal O}_j U_F\}$ can all be written in terms of the $\{ {\cal O}_j \}$, i.e. they generate the same commuting algebra.

\vspace{5pt}
An additional condition we can impose is that $U_F^N=1$ for some $N$.  We expect that these two conditions will lead to some sort of stability against heating, via an argument exploiting the many-body localizability of $U_F^N$ or the pre-thermalization ideas of Refs. \onlinecite{abanin2015exponentially, abanin2015effective,else2016pre} that were used in the time crystal context \cite{else2016floquet,von2016absolute}, but we leave this analysis to future work.

The key consequence of the compatibility property above is that Floquet evolution $U_F$ takes string operators $X$ to other string operators $U_F^\dag X U_F$.  Indeed, the defining property of a string operator in a model stably equivalent to the toric code is that it commute with all of the vertex and plaquette operators in the bulk of the string -- i.e. away from the endpoints -- as well as all of the ancilla $S^z$ spins (since it should not act on the latter in the bulk of the string).  This is just the requirement that the bulk of the string $X$ commute with the FSCLO, and since the constraint algebra generated by the FSCLO is invariant under $U_F$, this implies the same for $U_F^\dag X U_F$.

Now, the braiding properties of the topological excitations, encoded in the algebraic commutation properties of the string operators and preserved by local unitary conjugation, imply that an $e$ string operator can only map to an $e$ string operator or to an $m$ string operator.  In the later case we say that the original $U_F$ exchanges $e$ and $m$.  In appendix \ref{app:stable}, we give an example of a honeycomb Floquet unitary that exchanges $e$ and $m$.

\subsection{Chiral edge}


\begin{figure}[tb]
\begin{center}
\includegraphics[width=.95\linewidth]{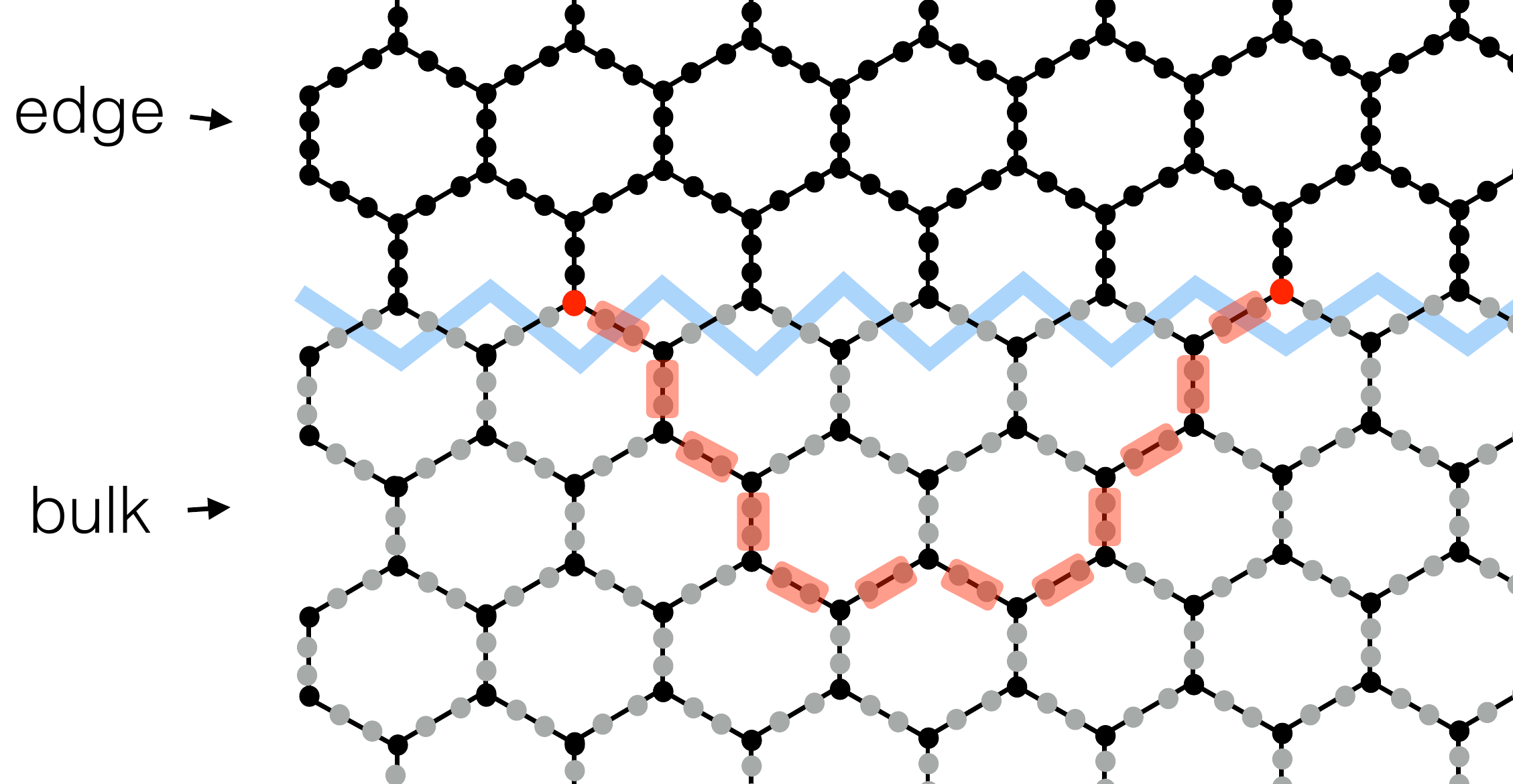}
\end{center}
\caption{Majorana representation of the honeycomb model.  The spins above the blue cut are part of the edge, and the ones below the blue cut are part of the bulk.  A fermionic string operator is indicated in red: its bulk portion contains the product of $\sigma_{r,r'}$ $\Z_2$ gauge field variables over a string.  Acting on the constrained Hilbert space where the bulk plaquette fluxes ${\cal F}_P$ take on prescribed values, it is equivalent to a string operator acting at the edge.  Here the open red rectangles run along a lattice $\Z_2$ gauge field defined at the edge.}
\label{fig:chiral_figure_3b}
\end{figure}

\begin{figure}[tb]
\begin{center}
\includegraphics[width=.95\linewidth]{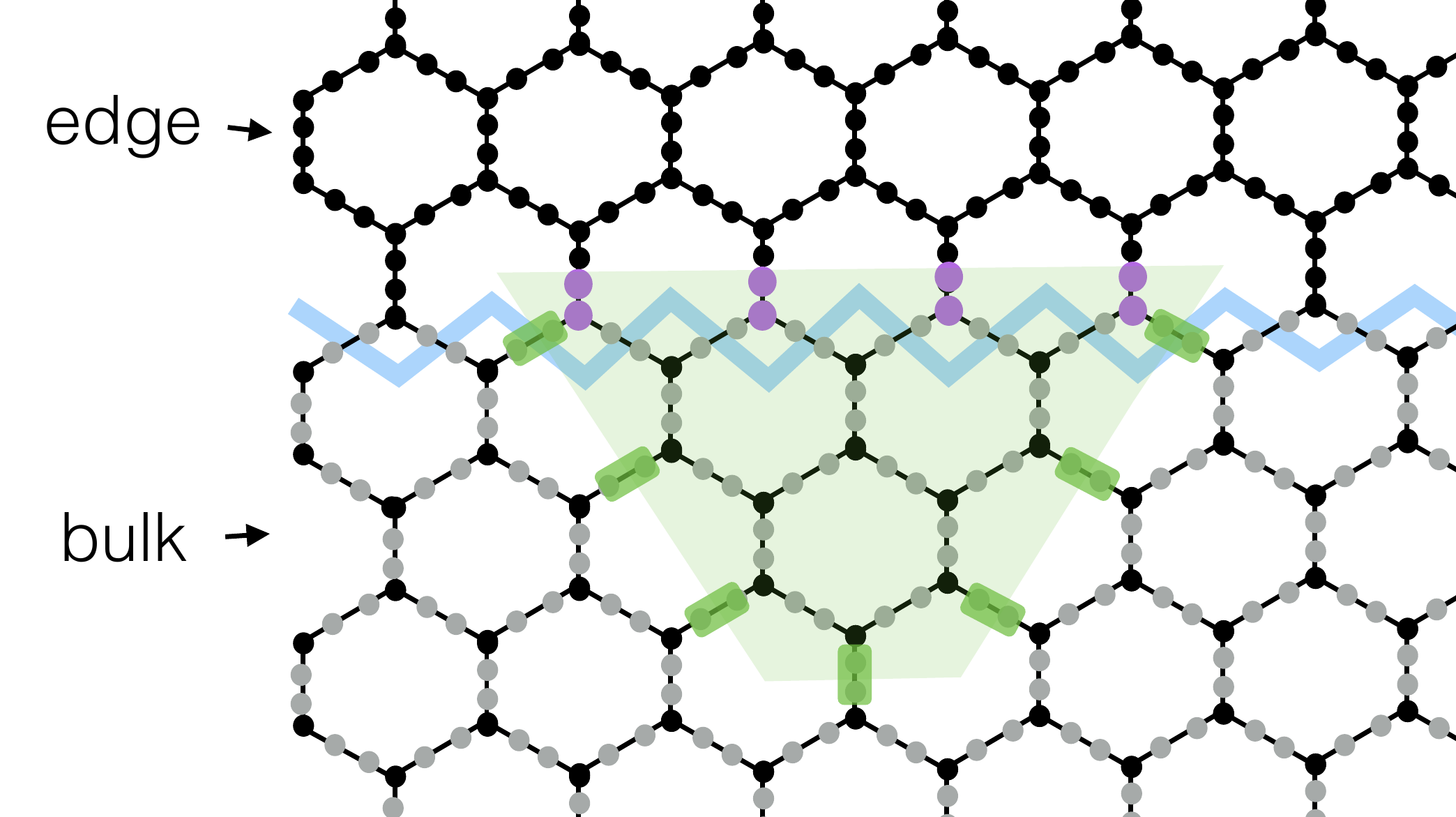}
\end{center}
\caption{Majorana representation of the honeycomb model.  A string operator that tunnels a $\Z_2$-vortex is indicated in green rectangles, corresponding to either an $e$ or $m$ string operator depending on the sublattice hexagons on which the string terminates.  By performing a global $\Z_2$ gauge transformation in the green region, and using the fact that in the constrained Hilbert space the ${\cal P}_{r,r'}$ quantities take on prescribed values in the bulk, we see that the action of this string operator on the constrained Hilbert space is equal to the product of the Majorana modes indicated in purple.  This product measures the fermion parity in the interval between the left and right tunneling endpoint.}
\label{fig:chiral_figure_3a}
\end{figure}

We now demonstrate how in a system with bulk toric code eigenstate order, enforcing the bulk conserved quantities as constraints yields a constrained Hilbert space that can be naturally interpreted as a Hilbert space of a quasi-1d fermionic system coupled to a $\Z_2$ gauge field.  By assumption, our system is stably equivalent to the toric code, and hence to the honeycomb model.  Recall that the honeycomb model \cite{kitaev2006anyons, FractionalCF} consists of spin-$1/2$'s $\vec{S}_r$ at sites $r$ of a honeycomb.  A useful representation of this Hilbert space is obtained by writing each spin-$1/2$ in terms of four Majorana mode variables $\{c_{r}, b_{r}^{x,y,z} \}$:

\begin{align}
S_{r}^i = i c_{r} b_{r}^i
\end{align}
together with the constraint $b_{r}^x b_{r}^y b_{r}^z c_{r}=1$.  Graphically, the $c_r$ operators are represented as sitting at the honeycomb sites, and the $b^i_r$ as sitting on the corresponding links, near the site.  $i=x,y,z$ corresponds to the three possible link orientations.  We also define $\Z_2$ gauge link variables $\sigma_{r,r'}=i b_{r}^j b_{r'}^j$, where $j \in {x,y,z}$ according to the type of link $\langle r,r' \rangle$, choosing an orientation so that each link $\langle r,r' \rangle$ is always oriented from one arbitrarily chosen sublattice towards the other.  As we show in appendix \ref{app:stable}, whatever the FSCLO of our original model, under the local unitary equivalence to the honeycomb model it must map to the canonical honeycomb model conserved quantities (plus possible ancilla $\sigma^z$ spins, which are not relevant to the subsequent discussion).  These are the flux operators ${\cal F}_P$, given by the product of $\sigma_{r,r'}$ along the links $\langle r, r' \rangle$ of a hexagonal plaquette $P$, together with fermion parity operators ${\cal P}_{r,r'} = i c_r \sigma_{r,r'} c_{r'}$ for the vertical links $\langle r,r' \rangle$.  In the fermionic representation it is easy to see that this is a full set of local commuting quantities: specifying all of the fluxes and all of the occupation numbers for the vertical-link pairing of Majorana modes specifies a state uniquely up to gauge equivalence.

Now introduce an edge into the system.  At some distance greater than all relevant Lieb-Robinson lengths away from the physical cut, the truncated Floquet unitary, unitarily transformed into the honeycomb model variables as in Eq. \ref{eq:trunc}, commutes with the honeycomb model ${\cal F}_P, {\cal P}_{r,r'}$ conserved quantities.  We consider all spins closer than this distance to the edge to be part of the effective quasi 1d edge, and the remaining spins to be part of the bulk -- see Fig. \ref{fig:chiral_figure_3b}.  The constrained Hilbert space is then defined by fixing (arbitrarily) the eigenvalues of ${\cal F}_P, {\cal P}_{r,r'}$ operators in the bulk.

The algebra of all operators that commute with the bulk constraints is given by operators that act only on the edge spins, together with string operators that can tunnel through the bulk.  Taking the quotient by the closed string operators in the bulk, which act as $\pm 1$ on the constrained Hilbert space, gives the operator algebra of the edge.  As is shown graphically and explained in Figs. \ref{fig:chiral_figure_3b} and \ref{fig:chiral_figure_3a}, this is simply the algebra of operators of a quasi 1d fermion coupled to a $\Z_2$ gauge field.  The idea is simply that one can use the fixed values of the bulk conserved quantities to deform the various string operators to the edge.  Under this correspondence, the fermion string operator maps to the product of two fermion parity odd operators connected by a $\Z_2$ gauge field string (Fig. \ref{fig:chiral_figure_3b}), whereas $e$ and $m$ string map, modulo endpoint operators, to operators which measure the fermion parity on the interval between the two tunneling endpoints.

\subsection{Bulk boundary correspondence}

Now consider a 2d Floquet operator $U_F$ that preserves the algebra generated by the bulk FSCLOs.  Note that this does not necessarily mean that all of these commuting local operators transform to precisely themselves times a phase under $U_F$; rather, all we demand is that the algebra generated by them is preserved under $U_F$.   Indeed, the fact that $U_F$ preserves the bulk FSCLO means that it maps string operators to string operators, and hence maps the operator algebra of the constrained edge defined above to itself.  As we saw above, this is just the operator algebra of a quasi 1d fermion coupled to a $\Z_2$ gauge field.

In order to apply our fermionic classification results, however, we have to extend this to an action on the full operator algebra of a fermionic quasi-1d system.  By picking a gauge we see that we already have an action on the set of all fermion parity ${\emph{even}}$ operators in this system, so all we have to do is to define an action on the odd operators.  Because fermionic string operators map to other nearby fermionic string operators, we see that an operator of the form $X_1 X_2$, with $X_1$ and $X_2$ spatially separated along the edge and fermion parity odd, must map to $X_1', X_2'$, where the $X_i'$ are fermion parity odd and supported near the corresponding $X_i$.  One is then tempted to say that $X_1$ should map to $X_1'$ and $X_2$ to $X_2'$.  However, this definition is ambiguous up to a phase, since only the overall phase of $X_1' X_2'$ is well defined.  However, if we demand that Hermitian operators map to Hermitian operators, the ambiguity is reduced to only a sign.  Furthermore, since the sign is fixed for any product of two odd local operators, this is a global sign ambiguity.  This global ambiguity cannot be fixed further: it corresponds to the fact that we can always modify the fermionic edge operator by multiplying it by the global edge fermion parity.  As an example, applying this construction to the honeycomb model of appendix \ref{app:stable}, we see that the fermionic edge automorphism is simply a Majorana translation, as illustrated in Fig. \ref{fig:stable_eq_fig4}.
 
We thus see that for any system stably equivalent to the toric code, a Floquet evolution that is compatible with the FSCLO (namely, preserves the commuting algebra generated by the FSCLO) induces a locality-preserving unitary $Y$ of a quasi-1d fermionic edge.  We are now in a position to apply our classification results for fermionic quasi-1d locality preserving unitaries.  Indeed, we claim that the 1d operator $Y$ is radical (i.e. has $\zeta=1$ in the above notation) if and only if the bulk Floquet evolution exchanges $e$ and $m$.  To see this, consider a string operator $X_e$ that tunnels an $e$ quasiparticle into the bulk at point $a$ and out at point $b$, as in figure \ref{fig:chiral_figure_3a}.  Then
\begin{align}
X'= X_e \left({U^F}^\dag X_e U_F\right)
\end{align}
is a string operator that either tunnels no topological charge (if $U_F$ fixes $e$ and $m$) or is a fermionic string operator (if $U_F$ exchanges $e$ and $m$).  But on the edge, according to the dictionary established above, the operator $X_e$ simply measures the total fermion parity $P_I$ of the interval $I=[a,b]$, so that $X'$ acts on the edge as $P_I \left(Y^\dag P_I Y \right)$.  According to Eq. \ref{eq:Teq}, this is nothing but $A_L A_R$, where $A_L$ and $A_R$ are local operators acting near $a$ and $b$ respectively, whose fermion parities diagnose the radical nature of $Y$.  Thus $Y$ is radical if and only if $X'$ is a fermionic string operator, which is the case precisely when $U_F$ exchanges $e$ and $m$ in the bulk.  This proves the bulk-boundary correspondence.

\section{Discussion}

In this work we constructed a many-body quantized invariant that classifies interacting two dimensional Floquet phases of fermions in the MBL setting.  The value of the invariant can be diagnosed by exposing an edge in the system and examining the action of the Floquet operator at the edge.  The signature of a non-trivial phase is that despite being locality-preserving, the edge evolution nevertheless cannot be generated by any truly 1d Floquet Hamiltonian.  We showed that such non-trivial 1d fermionic locality-preserving operators are completely classified by an index that takes values in either the rational numbers (the rational case) or rational  numbers times the square root of $2$ (the radical case).  The radical case is an intrinsically fermionic phenomenon, having no bosonic counterpart, and corresponds to a translation by a single Majorana mode.  Beyond the intricate construction of the many-body index, we also gave a simpler diagnostic for determining whether a given edge is radical or rational.

The Majorana SWAP model Hamiltonian explicitly breaks fermion number conservation, leaving only the fermion parity symmetry unbroken.  Although it can be realized as a mean field description of a paired state, the gapless Goldstone modes in a pair superfluid would cause problems for MBL, as would the long range interactions in a superconductor.  On the other hand, $\Z_2$ fermions can also be realized as an emergent description of a bosonic system with $\Z_2$ eigenstate topological order, and Ref. \onlinecite{FractionalCF} constructed a honeycomb model which had both an effective radical edge and had the bulk Floquet unitary acting as an anyonic symmetry, exchanging the $e$ and $m$ toric code excitations in the bulk.  In this case, at a finite density of $e$ and $m$ excitations the system behaves like a time crystal with a response at period $2T$.  In this work we used our classification of 1d locality preserving unitaries to give a general proof of this bulk-boundary correspondence.  In the course of this argument, we used the idea of stable equivalence to introduce a `strong' notion of eigenstate topological order.

On the other hand, there do exist models with equilibrium toric code topological order where the $e$ and $m$ excitations are exchanged by an onsite $\Z_2$ symmetry, which certainly cannot have an edge chirality\cite{Heinrich2016symmetry, meng2016exactly}.  The implication of the bulk-boundary correspondence is thus that, despite having ground state topological order, these models cannot have topological order in all eigenstates.  Indeed, these models are constructed using commuting projectors in such a way that the projectors are non-zero only if the state they act on satisfies some local constraints, leading to a massive degeneracy in their excited state spectrum.  The bulk-boundary correspondence implies that this degeneracy cannot be lifted, i.e. that it is impossible to construct a full set of toric-code like local integrals of motion compatible with the $\Z_2$-symmetry.  In particular, it implies that these models cannot be many-body localized by disordering the coefficients of their commuting projectors\cite{Chandran14,Bahri15}.

There are several avenues for extending the present work.  One is to relate the `strong' notion of eigenstate topological order introduced in this paper to other ways of characterizing eigenstate topological order, e.g. in terms of the existence of string operators commuting with the Hamiltonian.  Indeed, we expect that `strong' eigenstate topological order is actually equivalent to any other sufficiently precise definition of eigenstate topological order, but leave this issue for future investigation.

Another possible extension is to study more general topological orders, e.g. $\Z_n$ gauge theories.  In this case, a defect that exchanges the charge and flux excitations is known to bind parafermion zero modes \cite{lindner2012fractionalizing, clarke2013exotic,fendley2012parafermionic}, and it would be interesting to generalize the $\Z_2$-graded algebra based fermionic classification to a parafermionic one based on $\Z_N$-graded algebras.  We expect that this is possible, and will yield an extension of the many body index valued in rational numbers times $\sqrt{N}$.   Furthermore, it is natural to try to extend the present work to the case of models with additional global symmetries, such as $U(1)$ particle number conservation.  

\vspace{2mm}

{\bf{Acknowledgements -- }}   LF is supported by NSF DMR-1519579 and by Sloan FG-2015-65244.  AV acknowledges support from a Simons Investigator Award and AFOSR MURI grant FA9550-14-1-0035. This research was supported in part by the Kavli Institute of Theoretical Physics and the National Science Foundation under Grant No. NSF PHY11-25915.

\appendix

\section{Classification of 1d fermionic locality-preserving unitaries}\label{app:ferm1}

Our goal will be to extend the results of Ref. \onlinecite{gross2012index} to classify locality preserving unitary operators in fermionic Hilbert spaces.  We will follow the approach of Ref. \onlinecite{gross2012index} closely, generalizing their results on algebras to the case of $\Z_2$-graded algebras.

\subsection*{Notation and terminology}
For any set $S$, it will be useful to define $S_n$ as the enlargement of $S$ consisting of all sites at distance at most $n$ from some site in $S$.  We will then call an even unitary operator $U$ on $H$ `locality preserving' with range $n$ if, for any operator $X$ supported on a finite set $S$, $U^\dag X U$ is supported on $S_n$.

\subsection*{Step 1: Constructing the $\Z_2$-graded support algebras}

Suppose we have a locality-preserving unitary $U$.  The first step in our construction of the index of $U$ will be to define certain $\Z_2$-graded `support algebras' that characterize the quantum information flow associated to $U$.  To define these, we first coarse grain the Hilbert space by grouping sites in such a way that $U$ is locality-preserving with range $1$.  Now consider the operator algebra on two neighboring sites, $\cO_{2x} \otimes \cO_{2x+1}$.  We have that

\begin{align}\label{long_eq}
U^\dag & \left(\cO_{2x} \otimes \cO_{2x+1} \right) U \\
&\subset  \left(\cO_{2x-1} \otimes \cO_{2x}\right) \otimes \left(\cO_{2x+1} \otimes \cO_{2x+2} \right)
\end{align}
We now want to quantify the extent to which $U^\dag \left(\cO_{2x} \otimes \cO_{2x+1} \right) U$ is supported on either of the two tensor factors in brackets on the right hand side of the above equation.  To do this, we need to introduce the notion of a $\Z_2$-graded `support algebra'.

\vspace{2mm}

{\bf $\Z_2$-graded support algebra:} Let $\cB_1$ and $\cB_2$ be $\Z_2$-graded algebras of all linear operators on finite dimensional $\Z_2$-graded Hilbert spaces $H_1, H_2$ respectively, and let $\cA \subset \cB_1 \otimes \cB_2$ be some $\Z_2$-graded sub-algebra, closed under the taking of adjoints, i.e. under Hermitian conjugation.  Pick bases -- i.e. complete linearly independent sets of operators -- $\{ E_{\mu}^i \}$ of $\cB_2^i$ $(i=0,1)$, the even and odd parts of $\cB_2$.  Here $\mu$ ranges from $1$ to $|H_1^0|^2 + |H_1^1|^2$ for $i=0$, and from $1$ to $2|H_1^0| |H_1^1|$ for $i=1$.  Similarly, pick bases $\{ A_{\nu}^i \}$ of $\cA^i$.

Then any $A_\nu^i \in \cA^i$ has a unique expansion 
\begin{align}\label{expansions}
A_\nu^i &= \sum_\mu A_{\nu\mu}^{i0} \otimes E_\mu^0 + \sum_{\mu} A_{\nu\mu}^{i1} \otimes E_{\mu}^1
\end{align}
The algebra generated by all of the $A_{\nu\mu}^{ij} \in \cB_1$ is denoted $S(\cA,\cB_1)$ and called the support algebra of $\cA$ in $\cB_1$.  Since each $A_{\nu\mu}^{ij} \in \cB_1$ has well defined fermion parity $i+j$, $S(\cA,\cB_1)$ is a $\Z_2$-graded algebra.  Clearly, it has the property that $\cA \subset S(\cA,\cB_1) \otimes \cB_2$.  We claim that it is also the smallest $\Z_2$-graded algebra with this property, i.e. that for any other $\Z_2$-graded subalgebra $\cC \subset \cB_1$ for which $\cA \subset \cC \otimes \cB_2$, we must have $S(\cA,\cB_1)\subset \cC$.  Indeed, this just follows from the fact that we can expand each $A_\nu^i$ uniquely as a sum of operators in $\cC$ tensored with the $E_\mu^j$, showing that all of the $A_{\nu\mu}^{ij}$ are in $\cC$.  In particular, this shows that our definition of $S(\cA,\cB_1)$ is independent of the choice of bases taken above.  Furthermore, since $\cA$ was closed under the taking of adjoints, $S(\cA,\cB_1)$ must also have this property.

\vspace{2mm}

We now construct the support algebras:
\begin{align}
\cR_{2x}&=S\left(U^\dag(\cO_{2x} \otimes \cO_{2x+1}) U,\, \cO_{2x-1} \otimes \cO_{2x} \right)\\
\cR_{2x+1}&=S\left(U^\dag(\cO_{2x} \otimes \cO_{2x+1}) U,\, \cO_{2x+1} \otimes \cO_{2x+2} \right)
\end{align}

Let us examine some of the properties of the $\cR_y$.  The most important of these is that they all graded-commute.  Indeed, taking $\cR_{2x+1}$ for example, it is immediate that it graded-commutes with all $\cR_z$, except possibly $\cR_{2x+2}$.  To see that $\cR_{2x+1}$ and $\cR_{2x+2}$ graded-commute, we appeal to the following general result, the $\Z_2$-graded analogue of Lemma 8 in Sec. 7 of Ref. \onlinecite{gross2012index}: 

\vspace{2mm}

{\bf Claim: } Suppose $\cA \subset \cB_1 \otimes \cB_2$ and $\cA' \subset \cB_2 \otimes \cB_3$ graded-commute in $\cB_1 \otimes \cB_2 \otimes \cB_3$.  Then $S(\cA,\cB_2)$ and $S(\cA',\cB_2)$ graded-commute in $\cB_2$.  

{\bf Proof: } We use the following general fact:
\begin{align} \label{triple_identity}
[E\otimes X \otimes {\mathbb 1}, {\mathbb 1} \otimes X' \otimes E']_g = E \otimes [X,X']_g \otimes E'
\end{align}
for all $E\in \cB_1$, $X,X' \in \cB_2$, $E' \in \cB_3$.  This can be checked directly on elements $E, X, X', E'$ with well defined fermion parity, and then extended to all elements by linearity.

Now pick bases $\{E_\mu \}$ of $\cB_1$ and $\{E'_\nu \}$ of $\cB_2$, with each $E_\mu, E'_\nu$ having well defined fermion parity.  Take any $A \in \cA, A' \in \cA'$, and expand

\begin{align}
A &= \sum_\mu E_\mu \otimes A_\mu \\
A' &= \sum_\nu A'_\nu \otimes E'_\nu
\end{align}
We then have, using Eq. \ref{triple_identity},
\begin{align}
0 &= [A,A']_g \\
&= \sum_{\mu,\nu} E_\mu \otimes [A_\mu,A'_\nu]_g \otimes E'_\nu
\end{align}
Using the linear independence of the $\{E_\mu \}$ and of the $\{E'_\nu \}$, this implies that $[A_\mu,A'_\nu]_g=0$ for all $\mu,\nu$.  Since the set of all such $A_\mu$ and $A'_\nu$ generate $S(\cA,\cB_2)$ and $S(\cA',\cB_2)$ respectively, as we range over all $A \in \cA, A' \in \cA'$, we conclude that $[S(\cA,\cB_2),S(\cA',\cB_2)]_g=0$, as desired.

\vspace{2mm}

Applying the above claim to $\cA=U^\dag (\cO_{2x} \otimes \cO_{2x+1} ) U$, $\cA'=U^\dag (\cO_{2x+2} \otimes \cO_{2x+3}) U$, which clearly graded-commute and are contained in $\cB_1 \otimes \cB_2$ and $\cB_2 \otimes \cB_3$ respectively, with $\cB_j=\cO_{2x-3+2j} \otimes \cO_{2x-2+2j}$ for $j=1,2,3$, we see that $\cR_{2x+1}=S(\cA, \cB_2)$ and $\cR_{2x+2}=S(\cA', \cB_2)$ graded-commute, as desired.

Another important property of the $\cR_y$ follows from the fact that taken together, they generate the entire operator algebra $\cO$.  Indeed, since $\cR_{2x} \otimes \cR_{2x+1}$ contains $U^\dag(\cO_{2x} \otimes \cO_{2x+1})U$, the algebra generated by all of the $\cR_y$ contains all of the $U^\dag(\cO_{2x} \otimes \cO_{2x+1})U$.  But, since $U$ is unitary, the latter generate all of $\cO$ as we range over all $x$.

The fact that the $\cR_y$ generate all of $\cO$ implies that each $\cR_y$ has trivial graded center: in other words any element $Y \in \cR_y$ that graded-commutes with all of $\cR_y$ must be a multiple of ${\mathbb 1}$.  Indeed, any such $Y$ would then graded-commute with all the $\cR_z$, and hence all of $\cO$, but since $\cO$ is a matrix algebra, this means that $Y$ would be a multiple of the identity.

\subsection*{Step 2: Characterizing the $\Z_2$-graded support algebras} 

The properties of the $\cR_y$ that we derived above allow us to derive strong constraints on their form, which will be used in the definition of the fermionic index.  It will be instructive to first recall the bosonic case covered in Ref. \onlinecite{gross2012index}, where the $\cR_y$ are ordinary algebras.  The fact that each $\cR_y$ is a sub-algebra of a matrix algebra and closed under the taking of adjoints implies that it is semisimple.  The fact that $\cR_y$ also has trivial center then implies, using Wedderburn's theorem, that $\cR_y = {\mathbb C} (d_y)$, the algebra of $r(y)$ by $r(y)$ complex matrices, where $r(y)$ is some integer.

In our present $\Z_2$-graded case, each $\cR_y$ is still a sub-algebra of a matrix algebra and closed under the taking of adjoints, and so is still semisimple when viewed as an ordinary algebra, forgetting the $\Z_2$-graded structure.  Furthermore, as shown in the previous subsection, it has trivial graded center.  $\Z_2$-graded algebras that satisfy these two conditions are called `central simple superalgebras' (see golem.ph.utexas.edu/category/ 2014/08/ the underscore tenfold underscore way underscore part underscore 3.html).  There turns out to be a generalization of the Wedderburn theorem, the super-Wedderburn theorem (see above link), that states that these must be of one of two forms:

\vspace{2mm}
{\bf 1)} $\cR_y =  {\mathbb C}(p|q)$, the $\Z_2$-graded algebra of matrix operators on ${\mathbb C}^{p|q}$.  This has dimension $|\cR_y| = (p+q)^2$ as a vector space over the complex numbers.

{\bf 2)} $\cR_y = \Cl_1(p|q)$, the $\Z_2$-graded algebra of matrix operators on ${\mathbb C}^{p|q}$ with matrix entries taking values in the Clifford algebra $\Cl_1$ over the complex numbers.  Recall that $\Cl_1={\mathbb C} \oplus {\mathbb C}$ is the $\Z_2$-graded algebra consisting of elements of the form $a+b\gamma$, where $\gamma$ is an odd Hermitian generator (i.e. a Majorana mode).  $\cR_y$ then has dimension $|\cR_y|= 2(p+q)^2$ as a vector space over the complex numbers.

\vspace{2mm}
These two cases are referred to as even and odd simple $\Z_2$-graded algebras respectively \cite{Fidkowski11}.  For example, the Clifford algebras $Cl_n$ over the complex numbers are even or odd according to the parity of $n$.

\subsection*{Definition of fermionic index}
Having characterized the support algebras as above, we can prove one more useful fact, namely that
\begin{align}\label{equality1}
U^\dag(\cO_{2x} \otimes \cO_{2x+1})U = \cR_{2x} \otimes \cR_{2x+1}.
\end{align}
Indeed, we already know that $U^\dag(\cO_{2x} \otimes \cO_{2x+1})U \subset \cR_{2x} \otimes \cR_{2x+1}$, so all we have to prove is that the inclusion is an equality.  If it were not, then we could find an element $Z \in \cR_{2x} \otimes \cR_{2x+1}$, not proportional to ${\mathbb 1}$, that would commute with all of $U^\dag(\cO_{2x} \otimes \cO_{2x+1})U$.  $Z$ would also commute with all of the other $U^\dag(\cO_{2x'} \otimes \cO_{2x'+1})U \subset \cR_{2x'} \otimes \cR_{2x'+1}$, and hence with all of $\cO$, which is impossible since $\cO$ is a matrix algebra.

Taking the dimensions of the left and right hand sides of Eq. \ref{equality1}, we obtain:

\begin{align}\label{equality2}
(p_{2x}+q_{2x})^2 (p_{2x+1} + q_{2x+1})^2 = |\cR_{2x}| |\cR_{2x+1}|
\end{align}
Also, because $\cR_{2x+1}$ and $\cR_{2x+2}$ graded-commute, together their tensor product spans a $\Z_2$-graded sub-algebra of size $|\cR_{2x+1}| |\cR_{2x+2}|$ inside $\cO_{2x+1} \otimes \cO_{2x+2}$.  This is again an inclusion of $\Z_2$-graded even algebras, and by the same argument as above this inclusion cannot be strict, i.e. must be an equality.  From this we get:
\begin{align}
|\cR_{2x+1}| |\cR_{2x+2}|=(p_{2x+1}+q_{2x+1})^2 (p_{2x+2} + q_{2x+2})^2.
\end{align}
These two equations now show that the quantity
\begin{align}\label{defind}
&\frac{\sqrt{|\cR_{2x}|}}{(p_{2x}+q_{2x})} = \frac{(p_{2x+1}+q_{2x+1})}{\sqrt{|\cR_{2x+1}|}} \equiv \indf (U)
\end{align}
is independent of $x$; we call it the fermionic index of $U$.  Taking its logarithm, we define the fermionic chiral unitary index as:
\begin{align}
\nu_f (U) \equiv \log \indf (U)
\end{align}

\subsection*{Properties of the fermionic index}

{\bf Explicit formula:}  The fermionic index $\nu_f$ can also be expressed in terms of an explicit formula, given in Eq. \ref{eq:defnu}.  The proof of Eq. \ref{eq:defnu} parallels that of Lemma 12, Proposition 13, and Lemma 14 in Sec. 7 of Ref. \onlinecite{gross2012index}.  First, we define the measure $\eta({\cal A},{\cal B})$ that describes the extent to which $\Z_2$-graded algebras ${\cal A}$ and ${\cal B}$ fail to graded-commute.  

For the case of the site Hilbert spaces being $2^n$ dimensional fermionic Fock spaces, we can let ${\cal A}$ and ${\cal B}$ be generated by $2 n_A$ and $2 n_B$ Majorana modes respectively.  Then form monomials of these to generate sets of $p_A = 2^{2n_A}$ and $p_B=2^{2n_B}$ orthonormal operators $e^a_i, i=1,\ldots, p_A$ and $e^b_j, j=1,\ldots,p_B$ spanning ${\cal A}$ and ${\cal B}$ respectively.  Then, as in the main text, we define
\begin{align}\label{eq:defeta1}
\eta({\cal A},{\cal B}) = 2^{- n_{\Lambda}} \sqrt{\sum_{i=1}^{p_A} \sum_{j=1}^{p_B}
|\text{Tr}_{\Lambda} \left({e^a_i}^\dag e^b_j \right)|^2}
\end{align}
where
\begin{align}
n_{\Lambda}=\sum_{x\in \Lambda} n_x
\end{align}
is the total number of fermionic modes in the whole system, and the trace in Eq. \ref{eq:defeta} is taken in the Hilbert space of the whole system.

For the case of general $\Z_2$-graded site Hilbert spaces, we parallel the definition in Sec. 7 of Ref. \onlinecite{gross2012index}; instead of a basis generated by monomials of Majorana modes, we can take a general orthonormal linearly independent set.  The only subtlety is the issue of normalization -- the inner product in the space of operators is given by ${\text{Tr}} (A^\dag B)$, and depends on the size of the Hilbert space where $A$ and $B$ act.  However, this can be resolved just as in the case of bosonic site Hilbert spaces -- the $\Z_2$-graded nature of the Hilbert spaces poses no essential complication.  The final formula for $\eta({\cal A},{\cal B})$ is analogous to that given in Eq. 19 of Ref. \onlinecite{po2016chiral}, but slightly more cumbersome and not particularly enlightening, because the sums must be broken up into seperate even and odd sector sums.

Then the fermionic analogues of Lemma 12, Proposition 13, and Lemma 14 follow by replacing commutators with $\Z_2$-graded commutators and matrix algebras with even simple $\Z_2$-graded algebras.  These results show that the quantity
\begin{align} \label{eq:defnu1}
\text{log} \left(\frac{\eta(Y^\dag {\cal A}_L Y, {\cal A}_R)}{\eta({\cal A}_L, Y^\dag {\cal A_R} Y)}\right)
\end{align}
is equal to $1$ on finite depth circuits.  Since, as we have seen, it is equal to $\sqrt{2}$ on the Majorana translation, by the completeness property discussed below, it must be equal to $\nu_f$.  This proves Eq. \ref{eq:defnu}.

\vspace{2mm}

{\bf Completeness of classification: } 
\vspace{2mm}

{\bf Claim: } If $\indf (U)=\indf(U')$, then $U$ and $U'$ are stably-equivalent.

{\bf Proof: } Our proof is a refinement of the argument in Theorem 9 of Section 7 of Ref. \onlinecite{gross2012index}.  First, we assume to have coarse-grained our Hilbert space so that $U$ and $U'$ are both locality preserving with range $1$.  Let $\cR_y$ denote the support algebras in the above construction for $U$, and $\cR'_y$ those for $U'$.  Since $\indf(U)=\indf(U')$, $\cR_y$ and $\cR'_y$ have the same dimension for all $y$.  Now, there are two cases: either $U$ and $U'$ are both radical or both rational.  We claim that one can always reduce to the case when they are both rational.  Indeed, if they are both radical, one can append two ancilla spinless fermion systems.  Since performing opposite Majorana translations in these wires constitutes a finite depth quantum circuit, all one has to show is that $U$ and $U'$, when tensored with this circuit, are stably equivalent.  But this follows from showing that $U$ and $U'$, when tensored with one single such wire are stably equivalent, and these are both rational.

Thus we have reduced to the case that $U$ and $U'$ are both rational.  Now, even though $\cR_y$ and $\cR'_y$ are both even simple $\Z_2$-graded algebras of the same dimension, they might not necessarily be isomorphic.  This is different from the bosonic case, where we have ordinary simple algebras, i.e. matrix algebras ${\mathbb C} (n)$, which are uniquely determined by their dimension $n^2$.  This difference is what complicates the fermionic case and requires the additional notion of stable equivalence.  Indeed, in the fermionic case, all we can conclude is that $\cR_y={\mathbb{C}}(r_y|s_y)$, $\cR'_y={\mathbb{C}}(r'_y|s'_y)$ with $r_y+s_y=r'_y+s'_y=n_y$.  We will now simply tensor with an ancilla system consisting of single spinless fermion wire, with site Hilbert spaces ${\mathbb C}^{1|1}$.  Then the support algebras of $U \otimes 1$ and $U' \otimes 1$, denoted $\tcR_y$ and $\tcR'_y$, are just graded tensor products:

\begin{align}
\tcR_y&= {\mathbb{C}}(r_y|s_y) \otimes {\mathbb{C}}(1|1)={\mathbb{C}}(n_y|n_y) \\
\tcR'_y&= {\mathbb{C}}(r'_y|s'_y) \otimes {\mathbb{C}}(1|1)={\mathbb{C}}(n_y|n_y)
\end{align}
Thus, by tensoring with the spinless fermion wire ancilla, we can make the corresponding support algebras isomorphic.  For simplicity, I will now drop the tilde notation, and simply assume $\cR_y$ and $\cR'_y$ are isomorphic.  The proof now proceeds as in the bosonic case: because $\cR_y$ and $\cR'_y$ are isomorphic for all $y$, there exists a unitary operator $V_{2x-1} \in \cO_{2x-1} \otimes \cO_{2x}$ such that $V^\dag_{2x-1} \cR_y V_{2x-1}=\cR'_y$ for $y=2x$ and $y=2x-1$.  Let

\begin{align}
V=\prod_x V_{2x-1}
\end{align}
Then, since $U$ maps the operator algebra $\cO_{2x} \otimes \cO_{2x+1}$ to $\cR_{2x} \otimes \cR_{2x+1}$, $V$ maps $\cR_{2x} \otimes \cR_{2x+1}$ to $\cR'_{2x} \otimes \cR'_{2x+1}$, and $U'$ maps $\cO_{2x} \otimes \cO_{2x+1}$ to $\cR'_{2x} \otimes \cR'_{2x+1}$, we see that $(U')^{-1} VU$ maps each $\cO_{2x} \otimes \cO_{2x+1}$ to itself.  Thus we have

\begin{align}
(U')^{-1} VU=\prod_x V'_{2x}
\end{align}
where $V'_{2x} \in \cO_{2x} \otimes \cO_{2x+1}$ are unitaries.  Letting $V'=\prod_x V'_{2x}$, we then get that

\begin{align}
U=V^{-1} U' V'
\end{align}
so that $U$ and $U'$ differ by stacking finite depth unitaries, as desired.

\subsection{Examples}

{\bf Majorana chain: } Let us take the Hilbert space of spinless fermions, with $H_i={\mathbb C}^{1|1}$ on each site.  Under coarse graining, we can only generate sites whose dimensions are powers of $2$.  Eq. \ref{equality1} and Eq. \ref{equality2} then imply that $|\cR_y|$ is then an integral power of $2$ for all $y$, and hence the index of any locality preserving $U$ must also be the square root of an integral power of $2$.

The algebra of operators $\cO_i = {\mathbb C}(1|1) = \Cl_2$ on each $H_i$ is simply that generated by two Majorana modes, which we will call $\gamma_{2i-1}$ and $\gamma_{2i}$.  The total operator algebra $\cO_i$ is then $\cO={\mathbb C}(2^{N-1}|2^{N-1}) = \Cl_{2N}$, and is generated by the Majorana modes $\gamma_1, \ldots, \gamma_{2N}$.

Now, let $R$ be the $2N$ by $2N$ matrix defined by $R_{i,i+1}=1$ for $i=1,\ldots, 2N-1$, $R_{2N,1}=-1$, and all other $R_{i,j}=0$.  Since $R$ is in $SO(2N)$, we can write it as $R=\exp(A)$, with $A$ real and anti-symmetric.  Then define:

\begin{align} \label{Maj}
U_{\rm{Maj}}=\exp [\frac{1}{4} \sum_{i,j} A_{i,j} \gamma_i \gamma_j],
\end{align}
$U_{\rm{Maj}}$ then implements a Majorana translation: $U_{\rm{Maj}}^\dag \gamma_i U_{\rm{Maj}} = \gamma_{i+1}$ for $i=1,\ldots, 2N-1$ and $U^\dag \gamma_{2N} U = - \gamma_{1}$.  Using Eq. \ref{defind}, we find $\indf(U_{\rm{Maj}})=1/\sqrt{2}$.

{\bf General case: } We claim that any locality preserving unitary in any fermionic system is stably equivalent to either a bosonic locality preserving unitary in a bosonic system, or to such a bosonic locality preserving unitary stacked on top of the Majorana translation constructed above.  Indeed, since such systems span out the set of possible values for $\indf$, this just follows from the completeness of the fermionic classification discussed above.





\subsection {Diagnosing radical locality preserving unitaries} 

In this section we describe a simple way to determine whether $U$ is rational or radical, which avoids the complicated computation of $\indf(U)$ described in Eq. \ref{defind}.  First, let $P_i \in \cO_i$ be the operator that measures fermion parity at site $i$.  For any set $S$, let

\begin{align}
P_S = \prod_{i \in S} P_i
\end{align}
be the fermion parity within $S$, and let $P$ be the total fermion parity operator of the system:

\begin{align}
P=\prod_{i=1, \ldots, N} P_i
\end{align}
Now, given a locality-preserving unitary $U$, which we assume to have been coarse-grained to have range 1, and an interval $I=[a,b]$, consider the operator $T_I(U)$ defined by

\begin{align} \label{deft}
T_I(U) = U^{-1} P_I U P_I
\end{align}
Note that since the total fermion parity $P$ commutes with $U$ and with $P_I$, we have
\begin{align} \label{complement}
T_I&(U) = P^2 T_I(U) = U^{-1} (P P_I) U (P P_I) \\
&= U^{-1} P_{\bar{I}} U P_{\bar{I}},
\end{align}
where $\bar{I}$ is the complement of $I$.  Now, let's take an operator $X$ supported away from $I_1$, i.e. supported at least 2 sites away from $I$.  Then $U^{\dag} X U$ is also supported away from $I$, so both $X$ and $U^{\dag} X U$ commute with $P_I$.  Using Eq. \ref{deft}, we then see that $X$ must commute with $T_I(U)$.  By virtue of Eq. \ref{complement}, the same is true of any operator $X$ supported away from $\bar{I}_1$, i.e. in the interior of $I$ at least one site away from the endpoints.

More formally, this means that conjugating by $T_I(U)$ takes $\cO_i$ to itself for all $i$ except possibly $i=a-1,a,b,b+1$.    
Thus, conjugating by $T_I(U)$ takes $\cO_{a-1} \otimes \cO_a$ to some sub-algebra of $\cO_{a-1} \otimes \cO_a \otimes \cO_b \otimes \cO_{b+1}$.  Let us assume that $I$ is longer than $2$ sites, i.e. $a>b+1$.  Then, since $T_I(U)$ is locality-preserving with range at most $2$, this sub-algebra can only be $\cO_{a-1} \otimes \cO_a$ itself, i.e. conjugating by $T_I(U)$ takes $\cO_{a-1} \otimes \cO_a$ to itself, and similarly for $\cO_{b} \otimes \cO_{b+1}$.  This means that $T_I(U)=T_I^L(U) T_I^R(U)$, with $T_I^L(U) \in \cO_{a-1} \otimes \cO_a$, $T_I^R(U) \in \cO_{b} \otimes \cO_{b+1}$ being some unitary operators.  These two operators can either both be even or both be odd.

We now claim that $T_I^L(U)$ and $T_I^R(U)$ are both odd precisely when $U$ is radical.  To see this, we note first that the parity of these two operators depends only on the stable-equivalence class of $U$, is clearly multiplicative under stacking, and is clearly even for all bosonic locality-preserving unitaries.  Using the fact that any locality preserving fermionic unitary is stably equivalent to either a bosonic one or a bosonic one stacked with a Majorana translation, all we have to show is that the parity is odd for the Majorana translation defined in Eq. \ref{Maj}.  But for this specific case, we see directly that conjugating by $T_I(U_{\rm{Maj}})$ negates $\gamma_{a-1}$ and $\gamma_b$, and fixes all of the other $\gamma_i$.  Thus $T_I(U_{\rm{Maj}}) = \gamma_{a-1} \gamma_b$ up to phase, so that, up to phase, $T_I^L(U_{\rm{Maj}}) = \gamma_{a-1}$ and $T_I^R(U_{\rm{Maj}}) = \gamma_b$ are both odd, as required.  

\section{A modified honeycomb model} \label{app:stable}

\begin{figure}[tb]
\begin{center}
\includegraphics[width=1\linewidth]{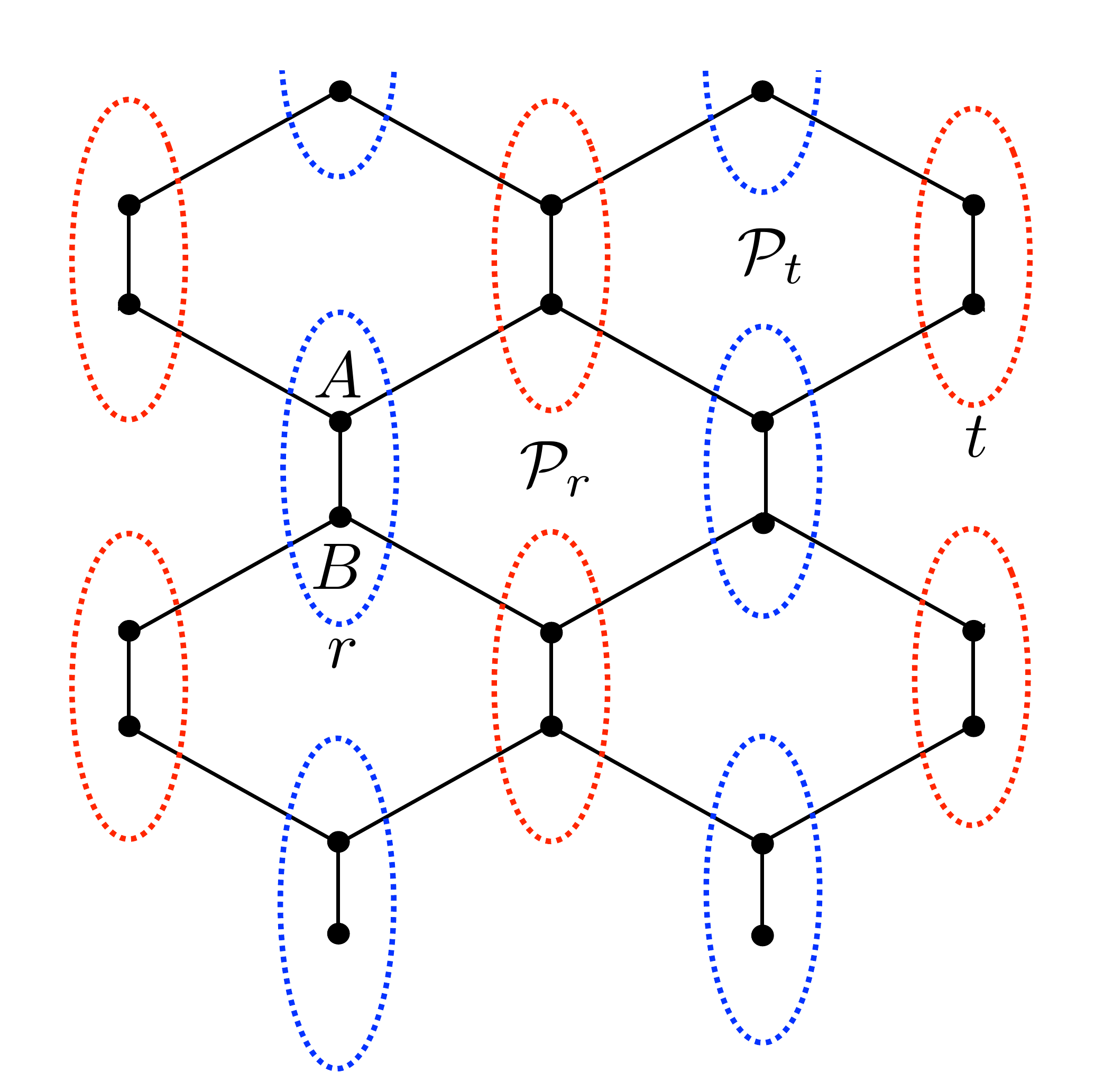}
\end{center}
\caption{Honeycomb model.  During the Floquet evolution, the observable ${\cal P}_t$ associated with red supersites picks up an Aharonov-Bohm phase ${\cal F}_{P_t}$ associated with the $\Z_2$-flux through plaquette $P_t$ located to the left of $t$.  Similarly, for blue supersites, the relevant plaquette is located to the right of $r$. }
\label{fig:stable_eq_fig1}
\end{figure}

\begin{figure}[tb]
\begin{center}
\includegraphics[width=.9\linewidth]{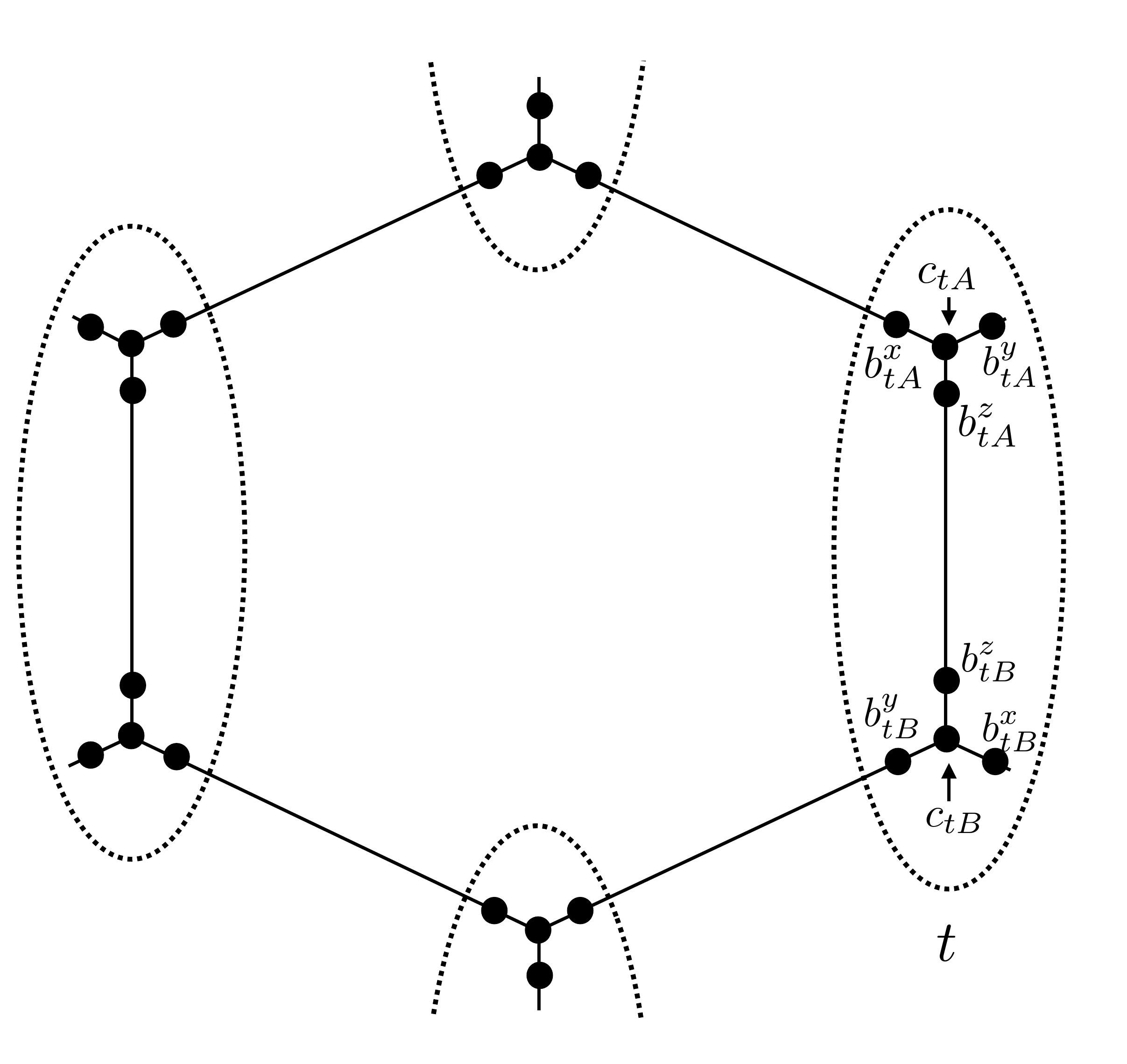}
\end{center}
\caption{Majorana representation of honeycomb model}
\label{fig:stable_eq_fig3}
\end{figure}

Let us define a slight variant of the honeycomb model of Ref. \onlinecite{FractionalCF}.  Our construction is based on Kitaev's honeycomb spin model, consisting of spin-$1/2$'s $\vec{S}_{tL}$, sitting on sites of a honeycomb.  Here $t$ denotes a supersite consisting of two vertically aligned nearest neighbor sites and $L=A,B$ is a sublattice index, as illustrated in Fig. \ref{fig:stable_eq_fig1}.  As in Ref. \onlinecite{kitaev2006anyons} it will be useful for us to represent this Hilbert space by writing each spin-$1/2$ in terms of four Majorana modes $\{c_{rL}, b_{rL}^{x,y,z} \}$:

\begin{align}
S_{rL}^i = i c_{rL} b_{rL}^i
\end{align}
We must impose the constraint $b_{rL}^x b_{rL}^y b_{rL}^z c_{rL}=1$ to reproduce a spin-$1/2$ Hilbert space.  Graphically, we represent the $c_{rL}$ as sitting at the honeycomb sites, and the $b^i_{rL}$ as sitting on the corresponding links nearby (see Fig. \ref{fig:stable_eq_fig3}).  We also define $\Z_2$ gauge link variables $\sigma_{rA,r'B}=i b_{rA}^j b_{r'B}^j$, where $j \in {x,y,z}$ according to the type of link $\langle rA,r'B \rangle$.  For definiteness we have taken the orientation to always go from the $A$ to the $B$ sublattice.  

\begin{figure}[tb]
\begin{center}
\includegraphics[width=.9\linewidth]{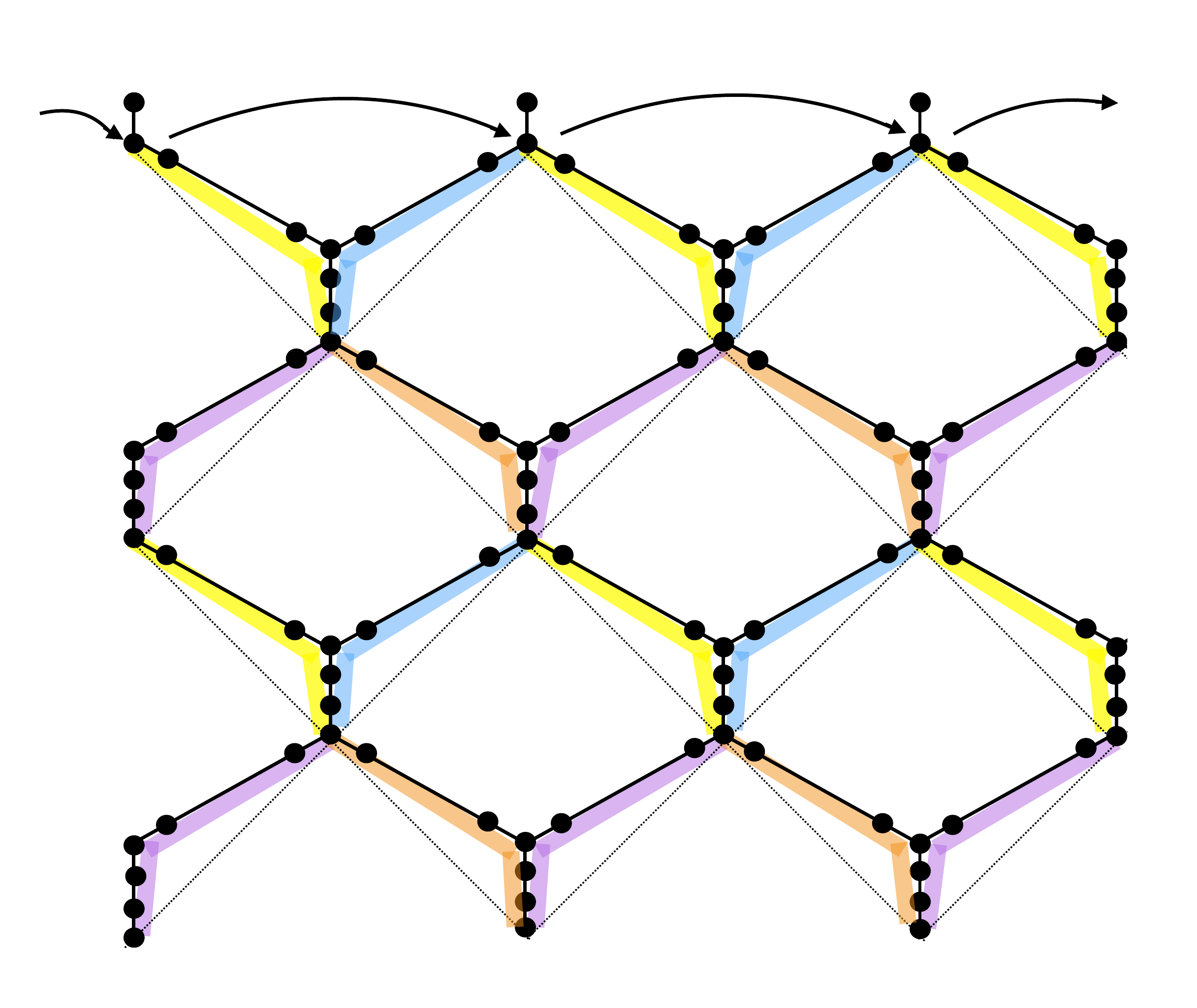}
\end{center}
\caption{Action of Floquet unitary on the edge}
\label{fig:stable_eq_fig4}
\end{figure}

Our Hamiltonian $H(t)$, periodic with period $T$, consists of $4$ piecewise constant driving terms $H_j$, $j=1,\ldots 4$, turned on for time $(j-1)T/4 \leq t < jT/4$, and is a slight variation on the one given in \onlinecite{FractionalCF}, in order to more directly relate it to the standard toric code below.  It is easiest to express in the fermionic variables.  The terms $H_j$ are each associated with hopping $B$-sublattice Majorana modes $c_{rB}$ between two nearest neighbor supersites, illustrated in figure \ref{fig:stable_eq_fig4} as yellow, blue, purple, and orange for $j=1,2,3,4$ respectively:

\begin{align}
H_j = \sum_{(u,t)\in j} \pi \frac{J}{4} \, i c_{uB} \left(\sigma_{tA,uB} \,\sigma_{tA, tB}\right) c_{tB},
\end{align}
where $(u,t) \in j$ means a pair of nearest neighbor supersites of the color associated to $j$.  If we fix all the $\Z_2$ gauge field variables to be equal to $1$, this is just the Majorana SWAP model.

Let us now analyze the resulting Floquet unitary 

\begin{align} \label{eq:defUF}
U_F(T) = T \exp \left( i \int_0^T dt \, H(t) \right)  
\end{align}
First, note that it leaves the gauge flux
\begin{align}
{\cal F}_P= \prod_{(rA,r'B) \in \partial P} \sigma_{rA,r'B}
\end{align}
through each hexagonal plaquette $P$ unaffected:
\begin{align}
U_F(T)^\dag {\cal F}_P U_F(T) = {\cal F}_P.
\end{align}
Now let
\begin{align}
{\cal P}_t = S_{tA} S_{tB} = c_{tA} \,\sigma_{tA,tB} \,c_{tB}
\end{align}
During the course of the Floquet evolution, the Majorana mode $c_{tB}$ hops around a plaquette $P_t$, located either to the left or to the right of $t$ as illustrated in Fig. \ref{fig:stable_eq_fig1}, and picks up an associated Aharonov-Bohm phase:

\begin{align}
U_F(T)^\dag {\cal P}_t U_F(T) = {\cal P}_t {\cal F}_{P_t}
\end{align}
This means that if there is a $\Z_2$ flux through plaquette $P_t$, then ${\cal P}_t$ changes sign.  Since ${\cal P}_t$ can be interpreted as fermion parity, this means that fermion parity changes in the presence of a $\Z_2$ flux, and hence $e$ and $m$ excitations get exchanged, as argued in \onlinecite{FractionalCF}.  Furthermore, as illustrated in Fig. \ref{fig:stable_eq_fig4}, the action on the edge consists of a Majorana translation.  Just as in \onlinecite{FractionalCF}, one can show that $U(2T)$ has an edge with well defined chiral unitary index equal to $\text{log} \, 2$, implying a fractional index of $\frac{1}{2} \text{log} \, 2$ -- for more details, see \onlinecite{FractionalCF, po2016chiral}.

\subsubsection*{Full set of commuting local operators for the honeycomb model}

\begin{figure}[tb]
\begin{center}
\includegraphics[width=1\linewidth]{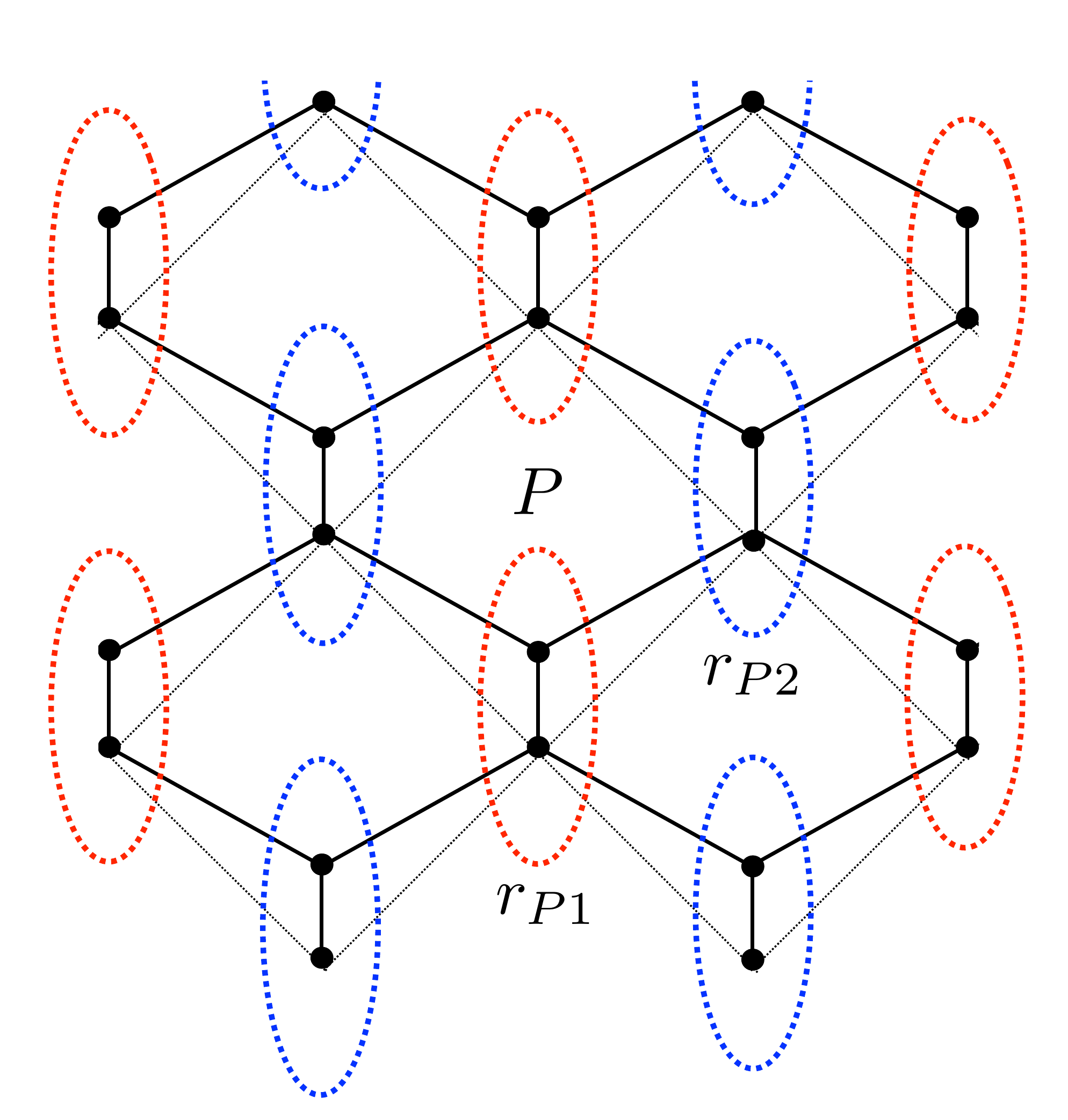}
\end{center}
\caption{Honeycomb model}
\label{fig:stable_eq_fig1_old}
\end{figure}

\begin{figure}[tb]
\begin{center}
\includegraphics[width=1\linewidth]{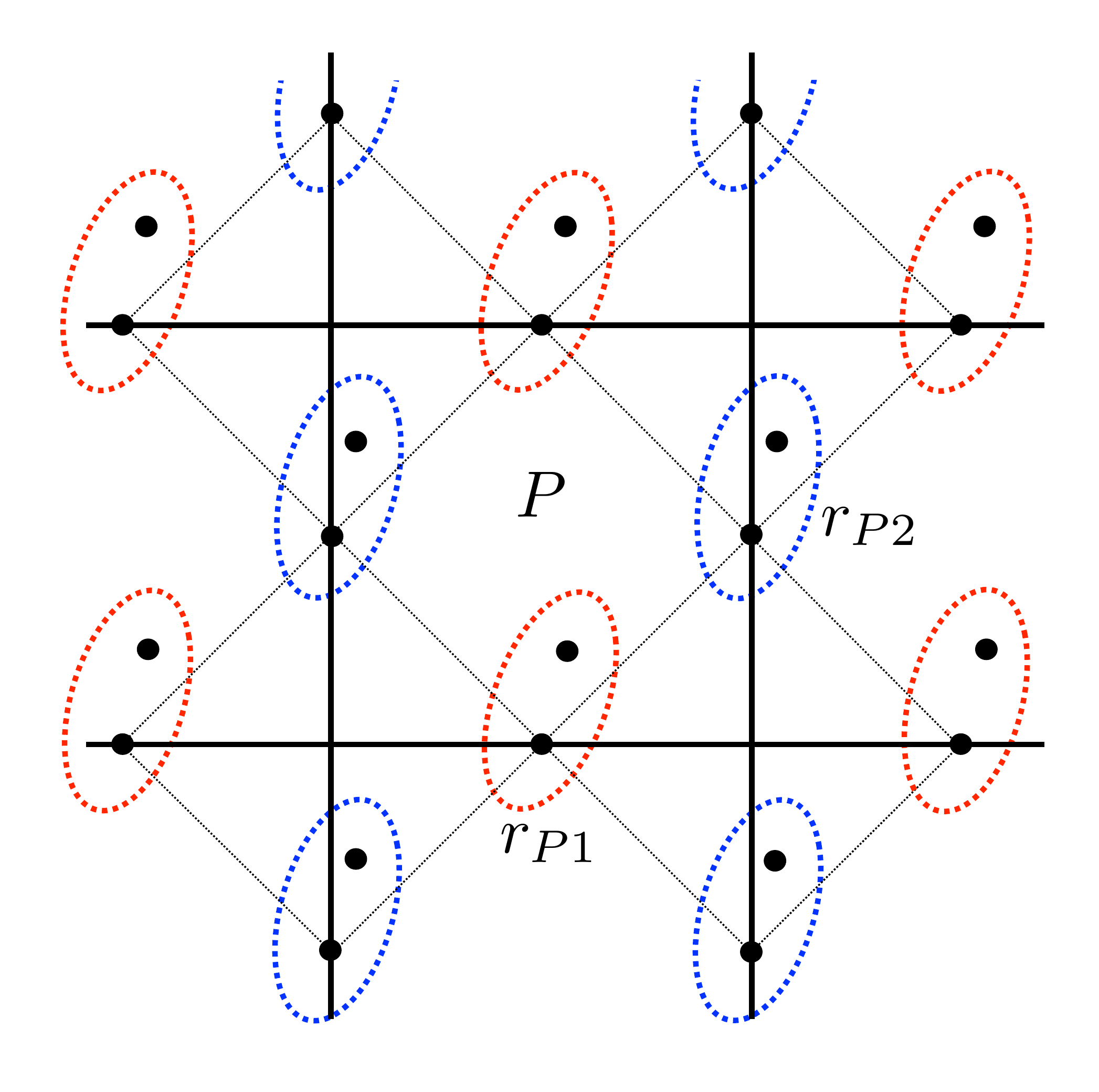}
\end{center}
\caption{Toric code with additional ancilla spins}
\label{fig:stable_eq_fig2}
\end{figure}

Let us now take the following FSCLO in the honeycomb model: $\{ {\cal F}'_P, {\cal P}_r \}$.  Here ${\cal F}'_P$ is a dressed version of the plaquette flux operator, defined as:

\begin{align}
{\cal F}'_P = {\cal F}_P {\cal P}_{r_{P1}} {\cal P}_{r_{P2}}
\end{align}
where $r_{P1}$ and $r_{P2}$ are the lower and right supersites of ${\cal P}$ respectively, as illustrated in Figs. \ref{fig:stable_eq_fig1_old} and \ref{fig:stable_eq_fig2}.  This is equivalent to the original FSCLO $\{ {\cal F}_P, {\cal P}_r \}$ in that we can recover each conserved quantity in one set as a product of conserved quantites in the other set.

Now consider a completely different system, namely the toric code on the square lattice.  We can take a FSLCO for the toric code consists of the standard vertex and plaquette terms ${\cal A}_V$ and ${\cal B}_F$, associated to vertices $V$ and faces $F$ of the toric code square lattice:

\begin{align}
{\cal A}_V &= \prod_{r \sim V} S_{rB}^x \\
{\cal B}_F &= \prod_{r \in \partial F} S_{rB}^z,
\end{align}
Here $r$ labels the links of the toric code lattice, and $B$ plays no role yet -- it is just an extra label. 

We now claim that the FSCLOs $\{ {\cal F}'_P, {\cal P}_r \}$ and $\{ {\cal A}_V, {\cal B}_F \}$ are stably equivalent.
To show this, we first have to establish an equivalence between the microscopic degrees of freedom in the toric code and the honeycomb model -- this is illustrated in figures \ref{fig:stable_eq_fig1_old} and \ref{fig:stable_eq_fig2}.  For the toric code, we take the Hilbert space consisting of a spin-$1/2$ on each link of a square lattice, and add an equal number of ancilla spin-$1/2$'s, as illustrated in Fig. \ref{fig:stable_eq_fig2}.  We will denote the operators associated to the spin on link $r$ by $S_{rB}^i$, and those associated to the corresponding ancilla spin by $S_{rA}^i$.  These microscopic degrees of freedom can be naturally identified with those of the honeycomb model, illustrated in Fig. \ref{fig:stable_eq_fig1_old}.  Furthermore, the honeycomb plaquettes can be naturally identified with those of the $45$ degree rotated $\frac{1}{\sqrt{2}}a$ lattice illustrated in Fig. \ref{fig:stable_eq_fig2}.  Half of these correspond to faces $F$ in the toric code square lattice -- we call these $P_F$ -- and the other half correspond to vertices of the toric code square lattice and are denoted $P_V$.

The FSLCO for the toric code consists of the standard vertex and plaquette terms ${\cal A}_V$ and ${\cal B}_F$, associated to vertices $V$ and faces $F$ of the toric code square lattice:

\begin{align}
{\cal A}_V &= \prod_{r \sim V} S_{rB}^x \\
{\cal B}_F &= \prod_{r \in \partial F} S_{rB}^z,
\end{align}
We claim that the FSCLOs $\{{\cal A}_V$, ${\cal B}_F\}$ and $\{ {\cal F}'_P, {\cal P}_r \}$ are stably equivalent.  We demonstrate this by explicitly defining a finite depth circuity unitary $U$, which actually turns out to be onsite, that takes one FSCLO to the other:

\begin{align}
U^\dag  {\cal F}'_{P_V} U &= {\cal A}_V \\
U^\dag {\cal F}'_{P_F} U &= {\cal B}_F \\
U^\dag {\cal P}_r U &= S_{rA}^z
\end{align}

Specifically, we define

\begin{align}
U = \left( \prod_{r \in \text{blue}} U_r \right) \left( \prod_{r \in \text{red}} U'_r \right)
\end{align}
where the colors refer to figure \ref{fig:stable_eq_fig1_old}, and where

\begin{align}
U_r = \exp \, \left[ i \frac{\pi}{4}(S^x_{rA} S^z_{rB} - S^x_{rA}) \right]
\end{align}
and
\begin{align}
U'_r  =& \exp \, \left[i \frac{\pi}{4} (S_{sA}^x-1)(S_{sB}^z-1)\right] \, \cdot \, 
\exp (-i \frac{\pi}{4} S^z_{sB}) \\ &\cdot \, \exp (-i \frac{\pi}{4} S^y_{sB}) 
\end{align}

The unitary that maps between the honeycomb model and toric code conserved quantities is defined by
\begin{align}
U = \left( \prod_{r \in \text{blue}} U_r \right) \left( \prod_{r \in \text{red}} U'_r \right)
\end{align}
where the colors refer to figure \ref{fig:stable_eq_fig1_old}, and where

\begin{align}
U_r = \exp \, \left[ i \frac{\pi}{4}(S^x_{rA} S^z_{rB} - S^x_{rA}) \right]
\end{align}
and
\begin{align}
U'_r  =& \exp \, \left[i \frac{\pi}{4} (S_{sA}^x-1)(S_{sB}^z-1)\right] \, \cdot \, 
\exp (-i \frac{\pi}{4} S^z_{sB}) \\ &\cdot \, \exp (-i \frac{\pi}{4} S^y_{sB}) 
\end{align}
To see that this is the case, let us first examine $U_r$.  It acts on the two spin-$1/2$ degrees of freedom in the $r$ vertical link of the honeycomb model as follows:
\begin{align}
U_r^\dag S^x_{rB} U_r &= i S^z_{rB} S^x_{rA} S^x_{rB}  \\
U_r^\dag S^z_{rB} U_r &= S^z_{rB} \\
U_r^\dag S^x_{rA} U_r &=S^x_{rA} \\
U_r^\dag S^z_{rA} U_r &= S^z_{rA} S^z_{rB}
\end{align}
This fully determines the action of $U_r$ on the operator algebra associated with the two spin-$1/2$'s.  Similarly, for the case of $U'_r$, we have:

\begin{align}
(U')^\dag_r S^x_{rB} U'_r &= S^z_{rB}  \\
(U')^\dag_r S^z_{rB} U'_r &=  i S^z_{rB} S^x_{rA} S^x_{rB} \\
(U')^\dag_r S^x_{rA} U'_r &=S^x_{rA} \\
(U')^\dag_r S^z_{rA} U'_r &= S^z_{rA} S^z_{rB}
\end{align}
Using these equations, we explicitly verify that:
\begin{align}
U^\dag  {\cal F}'_{P_V} U &= {\cal A}_V \\
U^\dag {\cal F}'_{P_F} U &= {\cal B}_F \\
U^\dag {\cal P}_r U &= S_{rA}^z
\end{align}
i.e. the conserved quantities of the honeycomb model map to those of the standard toric code, with ancilla spins added to the latter.

\subsubsection*{Floquet unitary exchanges $e$ and $m$ excitations}

\begin{figure}[tb]
\begin{center}
\includegraphics[width=1\linewidth]{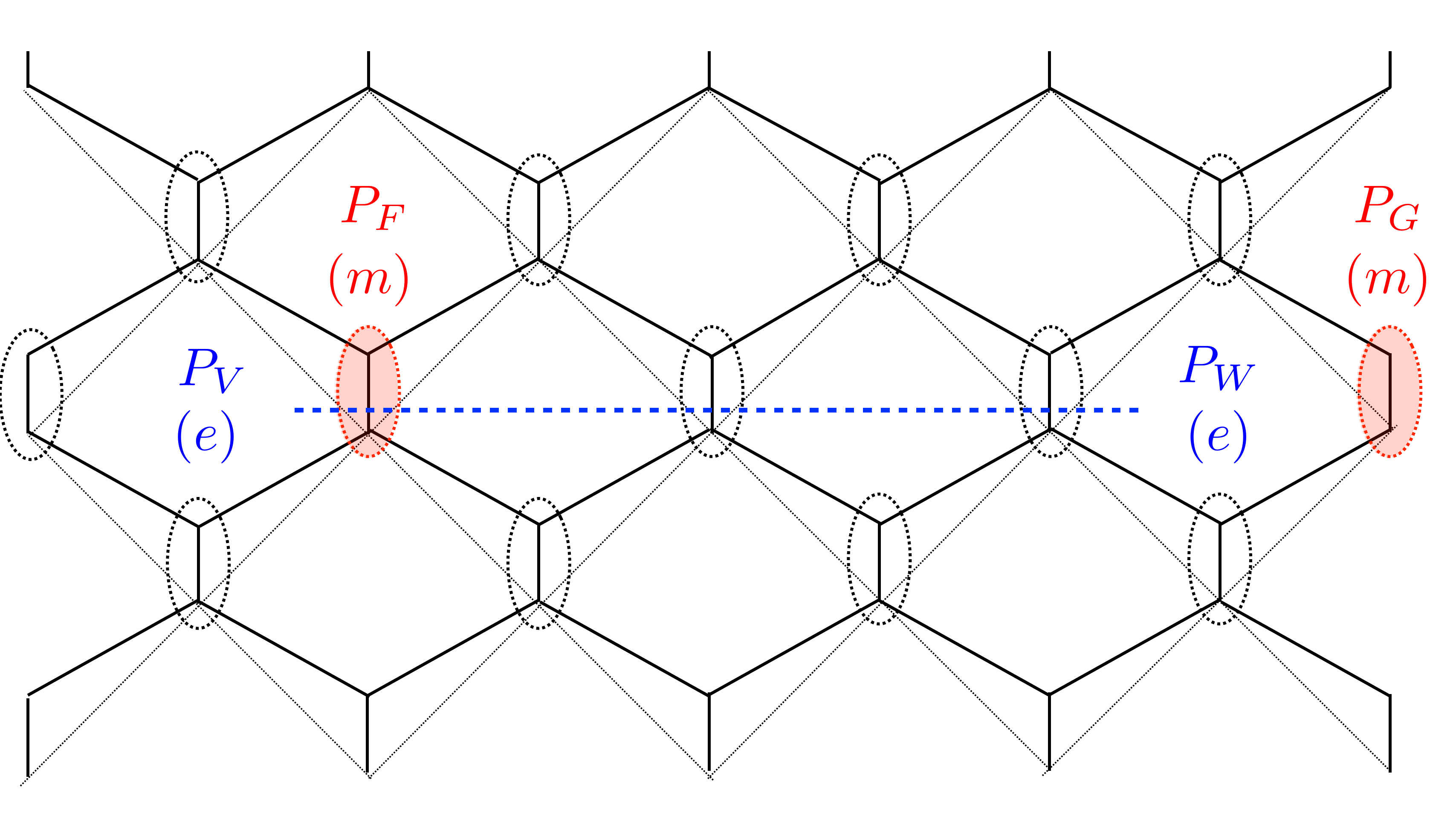}
\end{center}
\caption{The action of $U_F(T)$ turns an $e$ string operator into an $m$ string operator.}
\label{fig:chiral_figure_1}
\end{figure}

Let us see that the Floquet unitary $U_F(T)$ defined in Eq. \ref{eq:defUF} exchanges $e$ and $m$.  First, since it preserves the flux operators ${\cal F}_P$ and takes ${\cal P}_r$ to ${\cal P}_r {\cal F}_{P_r}$, it satisfies the compatibility condition of subsection \ref{subsec:eig}.  It can also be checked that $U_F(T)^4=1$, although we will not need this fact for the present analysis.

Now consider an $e$ string operator $X_e$ in our model.  This is just a string operator that creates $\Z_2$ flux excitations on widely separated $e$-type hexagonal plaquettes $P_V$ and $P_W$, as illustrated in Fig. \ref{fig:chiral_figure_1}.  Formally, it is an operator which near its left endpoint anti-commutes with ${\cal F}'_{P_V}$ and commutes with the rest of the FSCLO.  Now let 

\begin{align}
{\tilde X}_e = U_F(T)^\dag X_e U_F(T)
\end{align}
be the conjugated string operator.  Note that $U_F(T)$ and $U_F(T)^\dag$ flip the sign of any supersite operator ${\cal P}_r$ whenever the corresponding nearby plaquette $P_r$ has a non-trivial $\Z_2$ flux.  This means that all such supersite operators are flipped an even number of times by ${\tilde X}_e$, except those directly to the right of $P_V$ and $P_W$, where the $\Z_2$ flux changes between the application of $U_F(T)$ and $U_F(T)^\dag$.  Furthermore, all of the ${\cal F}_P$ operators are fixed by ${\tilde X}_e$.  Taken together, these facts imply that ${\tilde X}_e$ flips the sign of ${\cal F}'_P$ for $P=P_F, P_G$, as illustrated in Fig. \ref{fig:chiral_figure_1}.  Thus ${\tilde X}_e$ is an $m$ string operator (which also creates a local excitation corresponding to flipping the some ancilla spins near the string endpoints).

\bibliography{FloqFermion}
\end{document}